\documentclass[twocolumn]{aastex631}

\usepackage{eqnarray, amsmath}
\usepackage{mathtools}

\begin{document}

\title{Solar disk gamma-rays emission via synthetic magnetic field from photosphere to low corona}

\newcommand{\ep}{\textcolor{magenta}}

\author[0009-0009-5314-348X]{Eleonora Puzzoni}
\affiliation{Lunar and Planetary Laboratory, University of Arizona, 1629 E University Blvd, Tucson, AZ 85721, USA}
\affiliation{Observatoire de la Côte d’Azur, Laboratoire Lagrange, Bd de l’Observatoire, CS 34229, 06304 Nice cedex 4, France}

\author{Federico Fraschetti}
\affiliation{Lunar and Planetary Laboratory, University of Arizona, 1629 E University Blvd, Tucson, AZ 85721, USA}
\affiliation{Center for Astrophysics, Harvard \& Smithsonian, 60 Garden Street, Cambridge, MA 02138, USA}

\author{József Kóta}
\affiliation{Lunar and Planetary Laboratory, University of Arizona, 1629 E University Blvd, Tucson, AZ 85721, USA}

\author{Joe Giacalone}
\affiliation{Lunar and Planetary Laboratory, University of Arizona, 1629 E University Blvd, Tucson, AZ 85721, USA}



\begin{abstract}
Gamma-ray emission in the GeV-TeV range from the solar disk is likely to arise from  collisions of galactic cosmic rays (GCRs) with solar atmospheric plasma.
In a previous study, we demonstrated that closed turbulent magnetic arcades trap efficiently GCRs leading to a gamma-ray flux consistent with the Fermi-HAWC observations (from $\sim 0.1$ GeV to $\sim 1$ TeV).
Here, we model a synthetic magnetic field with a static, laminar structure of open field lines in the chromosphere increasingly braiding near the solar surface, with a scale height of $\sim 10^{-2} R_\odot$.
The height-dependent increase in magnetic field line braiding is modulated by an exponential scalar function, mimicking the bending of the photo- and chromo-spheric magnetic field revealed by polarimetric observations and reproduced by MHD simulations.
Employing 3D test-particle numerical simulations, we investigate how distorted magnetic field lines affect the gamma-rays production by injecting GeV-TeV protons into both magnetically laminar and braided regions. We find that with the chosen spatial resolution this synthetic magnetic field can account for the $> 10$ GeV gamma-ray spectrum observed by Fermi-LAT/HAWC. A  rebrightening between approximately $30$ and $100$ GeV (following a $\sim 30$ GeV spectral dip), suggests an enhanced confinement within the photo-/chromospheric layer by a stronger braiding.

\end{abstract}



\section{Introduction} 
\label{sec:intro}
Interest in high-energy gamma-ray emission from the solar disk has grown in recent years following the foundational work of \citet{Seckel1991}, hereafter SSG91, 
which introduced the first theoretical model based on gamma-ray production from GCRs mirroring into magnetic flux tubes and emitting gamma-rays in the interaction with ambient plasma of the solar atmosphere.
However, observations from Fermi-LAT \citep[LAT standing for Large Area Telescope, an imaging gamma-ray detector spanning energies from $\sim 20$ MeV to $\sim 300$ GeV, see][]{Atwood2009} revealed several limitations in the model: the predicted $\sim 1$ GeV flux was a factor of seven lower than the observations \citep[see][]{Abdo2011}. 
Secondly, while the SSG91 model predicted an abrupt cutoff at $\sim 5$ GeV, Fermi-LAT observed gamma-rays from the solar disk up to $100$ GeV \citep{Linden2018}, and the High-Altitude Water Cherenkov Gamma-ray Observatory (HAWC), which detects gamma-rays from $100$ GeV to $100$ TeV \citep{Abeysekara2013}, recorded gamma-rays up to $\sim 2.6$ TeV \citep{Albert2023}.
Other notable features in the solar disk gamma-ray emission include its anticorrelation with the solar cycle, initially detected by \cite{Ng2016} and subsequently confirmed by \cite{Linden2018, Linden2022} and \cite{Acharyya2025}. It was observed that the gamma-ray flux can be up to a factor of 2 lower during solar maximum compared to solar minimum.
\cite{Linden2022} (utilizing Fermi-LAT data spanning the entire 11-year solar cycle from 2008 to 2020) and \cite{Arsioli2024} (employing Fermi-LAT data from 2008 to 2022) both identified energy-dependent anisotropies in the gamma-ray emission from the solar disk.
In particular, \cite{Linden2022} found that emission at 10–50 GeV is nearly uniform across the solar disk, whereas above 50 GeV, it is concentrated along the equator with little polar emission. 
\cite{Arsioli2024} observed that during the solar maximum, higher-energy emission 
(20–150 GeV) is predominantly localized at the solar South pole, while lower-energy emission
(5–20 GeV) is concentrated at the North pole.
Additionally, the SSG91 model did not account for the observed spectral dip at $\sim 30$ GeV \citep[][]{Tang2018, Linden2018, Linden2022}, that was not detected by the CALorimetric Electron Telescope \citep[CALET,][]{Cannady2022}.

Several modeling attempts have been made to go beyond the SSG91 model and explain the new observations.
The solar atmospheric magnetic field is known to significantly influence gamma-ray emission \citep[see, e.g.,][]{Zhou2017}.
In order to address the gamma-ray excess from the disk, \cite{Li2024} proposed a two-component model with a flux tube (an open magnetic field line) and a flux sheet. 
Given the scarcity of studies focusing 
on closed photo-/chromospheric magnetic structures, \cite{Puzzoni2024} investigated the impact of closed magnetic arcades on gamma-ray emission from the solar disk. 
Smaller scale magnetic fluctuations were found essential to explain the $<30$ GeV 
Fermi-LAT flat spectrum. 
Additionally, \cite{Ng2024} estimated gamma-ray production semi-analytically by examining how the horizontal component of the magnetic field below the photosphere affects the flux in the TeV energy range (HAWC).
\cite{Alfaro2024} emphasized the significance of the photospheric magnetic field in modeling the GCR deficit, and the resulting excess of solar disk gamma-rays, at the Sun's location in the sky. This effect, known as the Sun shadow \citep[see, e.g.,][]{Amenomori2018, Becker2020}, varies with energy, as higher-energy GCRs are less deflected by the magnetic field and are therefore more likely to be blocked by the Sun, creating the shadow.
\cite{Gutierrez2022} analyzed the energy-dependent shadow of the Sun measured by HAWC to model gamma-rays, neutrinos, and neutrons excess.
\cite{Mazziotta2020} used FLUKA simulations to model the yields of secondary particles, including gamma-rays from the solar disk, produced by GCR interactions in the solar atmosphere.

The present study extends the analysis from \cite{Puzzoni2024} to a synthetic 3D model incorporating static open magnetic field lines that are increasingly distorted near the Sun's surface.
Specifically, our model captures the transition of the magnetic field near the Sun’s surface from predominantly radial 
to 
braided 
within the photosphere. 
This synthetic field is not derived from a physical driver, but is designed to mimic the braiding expected from processes such as granular flows, MHD turbulence, or magnetic reconnection, as shown by observations \citep[see, e.g.,][]{Morton2023}. In this sense, the model is conceptually aligned with 3D MHD-based reconstructions \citep[see, e.g.,][]{Hudson2020}.
The simulated low-photosphere field strengths range from $100$ G to $1$ kG, in agreement with both MHD models \citep[e.g.,][]{Shchukina2011, Rempel2014, delPino2018} and with observations \citep[e.g.,][]{Danilovic2010, Stenflo2013, Ishikawa2021}.
By adopting this model, we aim to explore in greater detail the migration of GCRs from the open magnetic field lines, which channel particles into the solar atmosphere, to the nearly-closed photospheric magnetic structures, an aspect that remained an open question in \cite{Puzzoni2024}. 
The particle trajectory is tracked with the PLUTO code \citep{Mignone2007} for distinct magnetic braiding. 

The paper is organized as follows. 
Section \ref{sec:model} outlines the magnetic field model. Section 
\ref{sec:numerical} details the numerical setup and analysis methodology. 
The results are presented in Section \ref{sec:results}, with 
Section \ref{sec:trend} focusing on how the GCRs interaction with ambient protons depends on the magnetic structure,
and Section \ref{sec:angular} examining the direction of outgoing gamma rays. Section \ref{sec:flux} compares the computed gamma-ray flux with Fermi-LAT and HAWC observations. Finally, Section \ref{sec:summary} summarizes the findings.

\section{Theoretical model}
\label{sec:model}

\subsection{Magnetic field structure}
\label{sec:turb_field}
\begin{figure}
    \centering
    \includegraphics[width=0.5\textwidth]{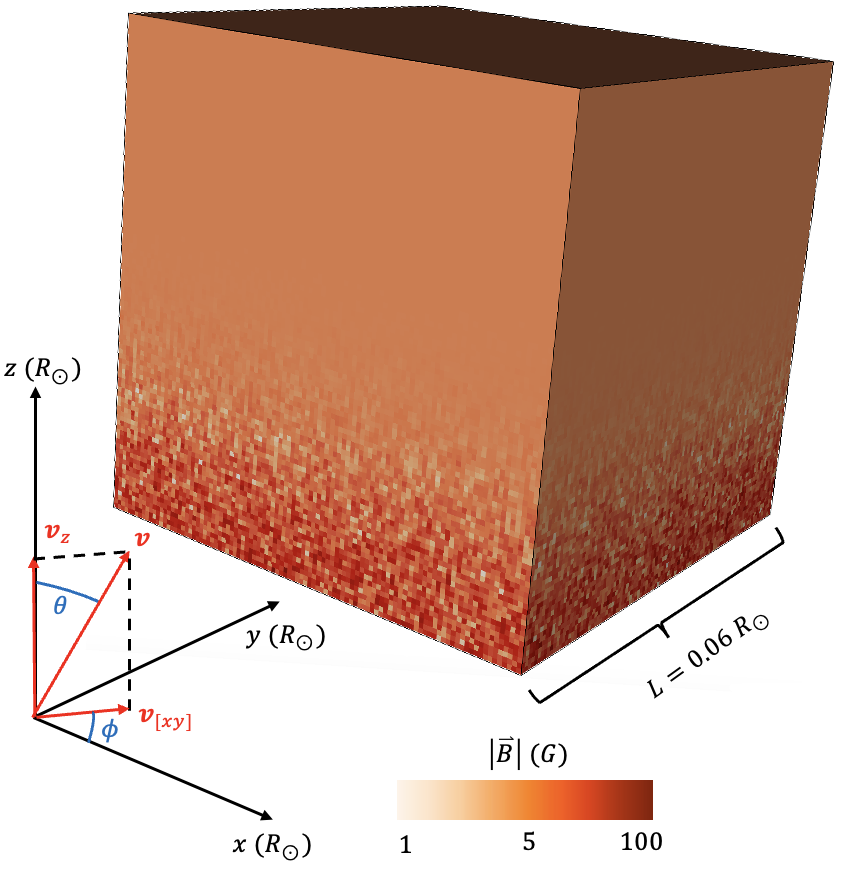}
    
    \caption{Magnetic field magnitude in the 3D domain (with logarithmic color scale in Gauss) together with the decomposition of the particle velocity vector defined by the angles $\theta$ and $\phi$. The xy-plane is tangent to the solar surface and z-axis is parallel to the local radial direction.}
    \label{fig:domain}
\end{figure}

The upper chromosphere/low corona magnetic field $B_0$ is assumed to be stationary on the particle motion time scale and uniform. We span the range $B_0 = 5- 50$ Gauss along the $z$-direction (see Fig.\ref{fig:domain} at $z \gtrsim 0.02 R_\odot$ up to the corona), consistent with recent spectropolarimetric measurements from the Daniel K. Inouye Solar Telescope \citep[DKIST; see][]{Schad2024} and with 3D-MHD resistive simulations of jets adjacent to arcades, or more generally closed structures \citep{Gonzales2018}. These values are also in agreement with the magnetic field strength at the top of the magnetic arcade as assumed in \cite{Puzzoni2024} and the mean field in \cite{Hudson2020}.
Such values of $B_0 = 5 - 50$ G in the low corona correspond to a photospheric magnetic field strength of $B_\mathrm{max} \sim 100-1000$ G, consistent with MHD simulations as well as remote sensing observations \citep[see, e.g.,][]{Danilovic2010, Stenflo2013}.

As GCRs approach the photosphere, the magnetic field acquires a 3D structure. Therefore, we use a synthetic 3D total magnetic field to mimick the complex 3D photospheric field that can be expressed as:
\begin{equation}
\label{eq:B}
    \mathbf{B}(x,y,z) = \mathrm{B_0} \mathbf{\hat{k}} + \nabla \times (f(z) \mathbf{S}(x,y,z)),
\end{equation}
satisfying the divergence-free condition $\nabla \cdot \mathbf{B}(x,y,z)$ = 0. 
Expanding Equation (\ref{eq:B}), we obtain
\begin{equation}
\label{eq:B2}
    \mathbf{B}(x,y,z) = \mathrm{B_0} \mathbf{\hat{k}} + f(z) \nabla \times \mathbf{S}(x,y,z) + \nabla f(z) \times \mathbf{S}(x,y,z),
\end{equation}
where $f(z)$ 
is an arbitrary scalar function modulating the onset of the transverse components, and $\mathbf{S}(x,y,z)$ represents the vector potential associated with the synthetic magnetic field $\delta\textbf{B}(x,y,z)$, i.e., $\delta \mathbf{B}(x,y,z) = \nabla \times \mathbf{S}(x,y,z)$, derived in \cite{Giacalone2021} as
\begin{equation}
    \mathbf{S}(x,y,z) = \sum_{n=1}^{N_m} \frac{A_n}{k_n} (\sin \alpha_n \hat{x}'_n + i \cos \alpha_n \hat{y}'_n) e^{i k_n z'_n + i \beta_n}.
    \label{eq:S}
\end{equation}
In our model, we take $f(z) = e^{ - (z/\Lambda)^2}$ (with $\Lambda \simeq 1.41 \times 10^{-2} R_\odot$ being the scale height).
The total magnetic field can thus be rewritten as
\begin{equation}
\label{eq:Btot}
    \mathbf{B}(x,y,z) = \mathrm{B_0} \mathbf{\hat{k}} + f(z)\delta \mathbf{B}(x,y,z) + \nabla f(z) \times \mathbf{S}(x,y,z).
\end{equation}
The spatial variation of $|{\bf B}|$ in the 3D domain of size $L^3$, where $L= 0.06 R_\odot$, is depicted in Figure \ref{fig:domain}: the magnetic field remains relatively uniform except for the low chromosphere and below, where distortions build up. 
The plane $z = 0$ represents the base of the photosphere, i.e., the surface where the vertical optical depth for radiation at a wavelength of $500$ nm reaches unity \citep{Athay:76}.
We note that $\mathbf{S}(x,y,z)$ and $\delta \mathbf{B}(x,y,z)$ have non-zero components along all three directions, while the third term on the rhs of Equation (\ref{eq:Btot}) is zero along the $z$-axis ($\partial_x f(z) = \partial_y f(z) = 0$). As a result, no component of the gradient-curvature drift velocity vanishes. 

$\delta \mathbf{B}(x,y,z)$ is defined in \cite{Giacalone1999} as:
\begin{equation}
    \delta \mathbf{B}(x,y,z) = \sum_{n=1}^{N_m} A_n (\cos \alpha_n \hat{x}'_n + i \sin \alpha_n \hat{y}'_n) e^{i k_n z'_n + i \beta_n},
\end{equation} with
\begin{gather}
 \begin{bmatrix} x' \\ y' \\ z' \end{bmatrix}
 =
  \begin{bmatrix}
   \cos \tilde{\theta}_n  \cos \tilde{\phi}_n & \cos \tilde{\theta}_n \sin \tilde{\phi}_n & -\sin \tilde{\theta}_n \\
   -\sin \tilde{\phi}_n & \cos \tilde{\phi}_n & 0 \\
   \sin \tilde{\theta}_n \cos \tilde{\phi}_n & \sin \tilde{\theta}_n \sin \tilde{\phi}_n & \cos \tilde{\theta}_n
   \end{bmatrix}
   \begin{bmatrix}
   x \\
   y \\
   z 
   \end{bmatrix},
\end{gather}
by summing over a large number of plane waves with wave mode $n$ (for a total of $N_m = 556$ modes), each characterized by amplitude $A(k_n)$, wavenumber $k_n$, randomly oriented polarization $\alpha_n$, and phase $\beta_n$. The angles $\tilde{\theta}_n$ and $\tilde{\phi}_n$ represent the random direction of propagation of each mode along the $z'$-axis of the local (primed) reference frame.
The random parameters are reshuffled every $100$ particles, leading to $100$ different realizations of the total magnetic field.
We adopt
\begin{equation}
  A^2(k_n) = \sigma'^2 G(k_n) \left[\sum_{n=1}^{N_m} G(k_n)\right]^{-1},
\end{equation}
where
\begin{equation}
    G(k_n) = \frac{\Delta V_n}{1 + (k_n L_s)^q}
\end{equation}
with $q = 11/3$. 
Here, $\sigma'^2 = \langle \delta\mathbf{B}^2(x,y,z)\rangle$, where $\langle \cdot \rangle$ indicates the average over the different magnetic field realizations, corresponds to the relative energy density of the magnetic components added to $B_0$, $\Delta V_n = 4 \pi k_n^2 \Delta k_n$ is the normalization factor dependent on the logarithmic spacing $\Delta k_n$ between $k_n$ (meaning that $\Delta k_n/k_n$ is a constant), and $L_s$ represents the largest horizontal scale of the magnetic photospheric structures. 

In this work, we set $L_s = 0.003$ $R_\odot \simeq 2$ Mm, comparable with the granular scale \citep[$\sim 1$ Mm, see, e.g.,][]{Trujillo2004}. $\mathbf{S}(x,y,z)$ in Eq. \ref{eq:S} spans a wavenumber interval from $L_\mathrm{min} = 10^{-3} R_\odot$ to  $L$,  corresponding to wavenumbers $k_\mathrm{max} = 2 \pi / L_\mathrm{min}$ and $k_\mathrm{min} = 2 \pi / L$, respectively. 
The assumed range in scales ensures resonant interactions across all GCR energies.
We explore different values of $\sigma^2 = \sigma'^2/B_0^2 = 0.1, 1, 10$.

\begin{figure*}
    \centering
\includegraphics[width=0.85\textwidth]{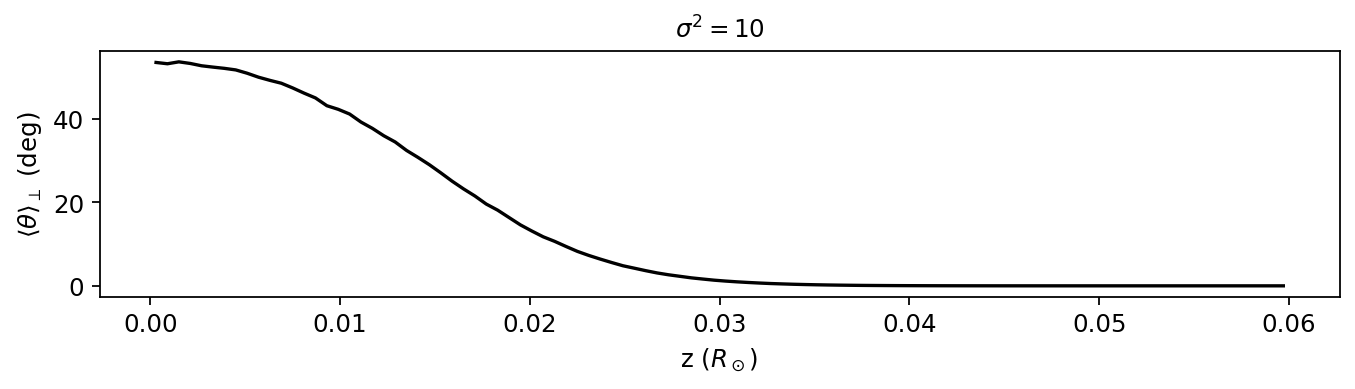}

    \includegraphics[width=0.49\textwidth]{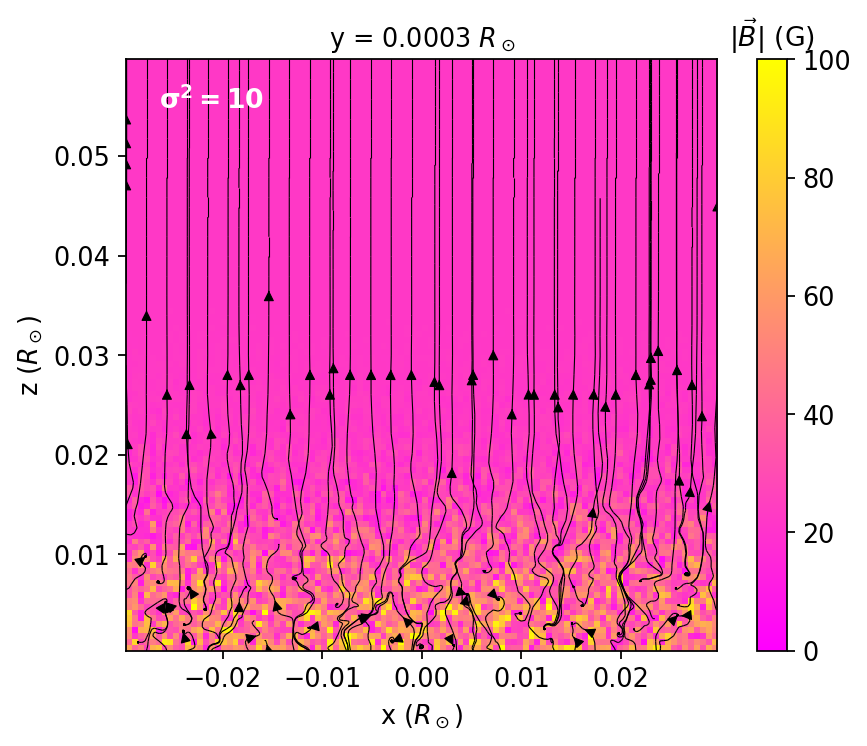}
    \includegraphics[width=0.49\textwidth]{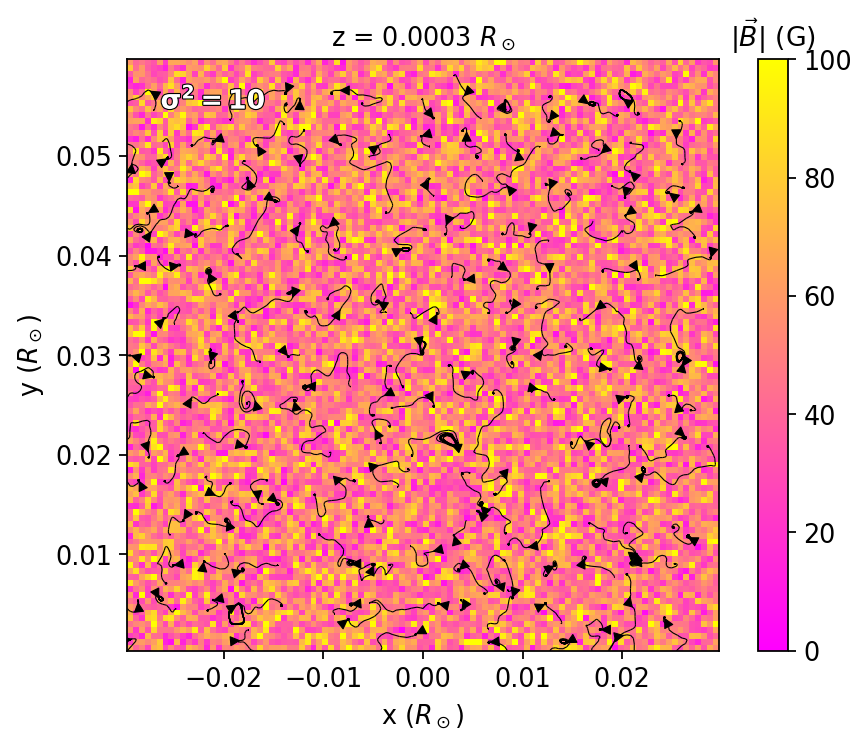}
    \caption{\textit{Top panel:} Average angle $\langle \theta \rangle_\perp$ (in degrees) as a function of the altitude $z$ with $\sigma^2 = 10$. \textit{Bottom panels:} Magnetic field lines (in black) and magnitude (colorbar) in the $xz$-plane (lowest $y = 0.0003 R_\odot$ plane in the grid, left panel) and $xy$-plane (lowest $z = 0.0003 R_\odot$ plane in the grid, right panel) with $\sigma^2 = 10$.}
    \label{fig:laminar.png}
\end{figure*}

The upper plot of Figure \ref{fig:laminar.png} shows the average angle $\langle \theta \rangle_\perp = \arctan(\sqrt{B_x^2(x,y,z) + B_y^2(x,y,z)}/B_0)$ between the magnetic field and the vertical direction $\hat{z}$ as a function of $z$, illustrating the deviation of the magnetic field from the vertical as it approaches the photosphere base. This definition of $\langle \theta \rangle_\perp$ is clearly an upper limit for $\sigma^2 \gg 1$, that includes longitudinal compressions. The $\langle \theta \rangle_\perp$ monotonically increases downward with a peak value for $\sigma^2 = 10$ at $50^\circ - 60^\circ$ from the vertical or $\sqrt{B_x^2 + B_y^2}/B_0 = 1.2 - 1.7$. This predominance of the horizontal component of the photospheric field is confirmed by high-resolution measurements, that report an average value $\sim 2.0 - 5.6$ \citep{Steiner.etal:08}. As a further clarification, this distortion is not the result of turbulent cascade or wave-wave collisions and does not contradict observations that identify maximal turbulent fluctuations at an intermediate height within the photosphere \citep[see, e.g.,][]{Schad2024}.
The general trend of vertical magnetic field dominance in the chromosphere and low corona, contrasted with horizontal field dominance in the photosphere, is also observed by \cite{Danilovic2010} and \cite{Stenflo2013}. We note that our synthetic model cannot reproduce the peak in $\langle \theta \rangle_\perp$ due to the small-scale dynamo in the photosphere found numerically by 
\cite{Rempel2014}, as the peak cannot be resolved by the grid resolution for the values of $B_0$ adopted here, comparable with the vertical field therein.
The magnetic field lines and magnitude 
are shown in the lower left ($xz$-plane) and right ($xy$-plane) panel of Figure \ref{fig:laminar.png} for a single realization of the total magnetic field with $B_0 = 5$ G and $\sigma^2 = 10$.
Note that $L_s < \Lambda$ to ensure that the largest magnetic structures are contained in the scale height $\Lambda$. 
The increasing distortion of the laminar field lines in the lower left panel is visible below $z < L/2= 0.03 R_\odot$, corresponding to the height of the closed arcade in \cite{Puzzoni2024}.
Small-scale magnetic structures form at $z \lesssim 0.01 R_\odot$, consistent with the heights of similar structures reported by \cite{Hudson2020}.
The lower right panel shows magnetic structures near the bottom of the photosphere, with sizes and a maximum magnetic field strength of $\sim 100$ Gauss with a large filling factor that aligns with the values inferred by combining radiative transfer simulations with polarization observations \citep{Trujillo2004}, that ruled out earlier estimates of the magnetic field at least two orders of magnitude smaller in the photosphere \citep[see, e.g.,][]{Faurobert-Scholl1995, Faurobert2001}. Here, $B_0 = 50$ G with $\sigma^2 = 10$ leads to a maximum magnetic field strength of $\sim 1$ kG, consistent with observations \citep[see, e.g.,][]{Stenflo2013}.

The synthetic field $\delta \mathbf{B} (x, y, z)$ is periodic in both the $x$- and $y$-directions. The wavenumbers $k_x$ and $k_y$ are selected to be close to randomly chosen values but must be multiples of $L$. It is essential that these new wavenumbers maintain consistency with the specified $k$:
if the new values of $k_x$ and $k_y$ lead to $k_x^2 + k_y^2 > k^2$, the values must be re-selected, using a new random number.

\subsection{Density profile}\label{sec:dens_prof}

Following \cite{Puzzoni2024}, we incorporate the density profile from 1D hydrostatic models \citep{Fontenla1993} for the low photosphere ($\rho_0 \sim 10^{-6}$ g/cm$^3$ at $z = 0$) up to $4$ Mm \citep{Morton2023} and from \cite{Gonzales2021} up to $30$ Mm.
The motion of GCRs in the total magnetic field $\mathbf{B}(x,y,z)$ in Eq. \ref{eq:B} is computed by solving the Lorentz equation.
In the upper chromosphere, the uniform magnetic field $|\mathbf{B}(x,y,z)| = B_0 = 5$ G, with a proton density $\rho \sim 10^{-16}$ g/cm$^3$ \citep[from][]{Gonzales2021}, results in a typical Alfvén speed $v_A = B_0/\sqrt{4 \pi \rho} \simeq 1.4 \times 10^8$ cm/s.
The ratio between this Alfvén velocity to the particle speed ($v \sim c$) is $v_A/c \simeq 4.7 \times 10^{-3}$, which validates the magnetostatic assumption also for the high fields and very dense low photosphere.

    

%
%

\section{Numerical setup}
\label{sec:numerical}

\subsection{Computational domain}
Test-particle protons 3D numerical simulations are conducted with the PLUTO code \citep[][]{Mignone2007, Mignone2012}.
We adopt a 3D computational domain with dimensions $L^3$. In particular, the domain is defined within $-0.03 < x/R_\odot < 0.03$, $0 < y/R_\odot < 0.06$, and $0 < z/R_\odot < 0.06$. The two-dimensional resolution of $1000^2$
used in \cite{Puzzoni2024} is extended to three dimensions in this work, employing a grid of $100^3$ cells with a spatial resolution of $6 \times 10^{-4} R_\odot \simeq 418$ km. 
As explained in the previous section, periodic boundary conditions are employed in the $x$- and $y$-directions.
Our simulations show that the GCR excursion out of the box sides before either GCR-proton interaction or GCR escape both downward into the Sun and upward toward the corona is only a few times $L$, thus $\simeq 0.1 - 0.2 R_\odot$,  supporting the approximation of the Sun surface with its tangent $xy$-plane.
Outflow boundary conditions are employed along the  $z$-direction for particles precipitating into the Sun surface or leaving the chromosphere with no interaction. 

\subsection{Particles initial conditions and motion in the magnetic field}
GCR protons are initialized with one particle per cell at two distinct heights: in the laminar (i.e., $z = z_\mathrm{up} = 0.05 R_\odot \sim 35000 \ km)$, and braided field region ($z = z_\mathrm{down} = 0.005 R_\odot \sim 3500 \ km$). 
In each simulation run, a total of $N_\mathrm{inj} = 10,000$ protons are injected. 
GCRs are injected at both heights with an isotropic velocity distribution as in \cite{Puzzoni2024}, supported by the observed isotropy of the anomalous cosmic rays (ACRs) up to a few MeV as close as $\sim 0.1$ AU \citep{Rankin2024}.
%
%

The particle orbits are integrated using the Boris 
algorithm, which is already implemented in the PLUTO code \citep{Mignone2007, Mignone2012}. The total magnetic field is 
interpolated \citep{Birdsall1991} at the particle's position \citep[see][for further details on the interpolation]{Mignone2018}.
Orbit integration continues until $t_\mathrm{stop} = 25 \ R_\odot/c \sim 58 \ s$, which corresponds to the time when all GCRs have either interacted within the computational domain or escaped from it. 

\subsection{Interacting protons and gamma-rays flux}
As in \cite{Puzzoni2024}, the interaction time $t_\mathrm{int}$ of the GCR proton-solar atmosphere proton is taken into account by 
\begin{equation}
    \Delta t / t_\mathrm{int} = \int_{t_i}^{t_f} dt/t_{int}(z) =   \sigma_\mathrm{pp} v \int_{t_i}^{t_f} n[z(t)] dt,
    \label{eq:t_int}
\end{equation}
where $\Delta t$ is the total elapsed time from $t = 0$ up to the final time $t_f$, $dt$ is the time-step of the code, $v \simeq c$ is the typical GCR speed, and $n[z(t)]$ represents the number density in the solar atmosphere of single proton (in the nearly fully-ionized chromosphere) or neutral Hydrogen (in the photosphere) .
The total inelastic cross section of the proton-proton interaction $\sigma_\mathrm{pp}$ is obtained from Equation 1 of \cite{Kafexhiu2014}.
We are interested in photons produced within the computational domain ($\Delta t/t_\mathrm{int} > 1$), and observed at Earth ($v_z > 0$).
Indeed, gamma-rays are assumed to be emitted along the same direction as the GCRs at $t=t_{int}$ due to momentum conservation (see Section \ref{sec:angular}).

The typical photon energy is $E_\gamma \sim 0.1 E_p$ via the $\pi^0$ mesons primary channel \citep[][]{Kelner2006} for proton-proton interactions \citep[i.e., $p + p \rightarrow \pi^0 \rightarrow \gamma + \gamma$, see][for a comparison with empirical hadronic interaction models]{Dorner2025}.
Following the method outlined in \cite{Puzzoni2024} and \cite{Li2024}, we calculate the gamma-ray flux:
\begin{equation}
\label{eq:flux*}
    \Phi^*_\gamma (E_\gamma) \equiv \frac{dN_\gamma}{dE_\gamma} = c n_p\int_{E_\gamma}^\infty \sigma_\mathrm{pp}(E_p) J_p(E_p) F_\gamma \left(\frac{E_\gamma}{E_p}, E_p \right)\frac{dE_p}{E_p},
\end{equation}
where $n_p$ is the proton number density integrated along the $z$-direction within the dense atmosphere, 
$F_\gamma \left(\frac{E_\gamma}{E_p}, E_p \right)$ is an empirical function accounting for the flux of outgoing $E_\gamma$ photon from an impinging $E_p$ proton \cite[][Eq. 58 therein]{Kelner2006}, and $J_p(E_p) = N_\mathrm{int}(E_p)/N_\mathrm{inj} \times (E_p^2 dN/dE_p)_\mathrm{obs} \times 1/(E_p^2 c)$,
where $N_\mathrm{int}/N_\mathrm{inj}$ represents the ratio of interacting to injected protons, averaged over the injection altitudes $z_\mathrm{up}$ and $z_\mathrm{down}$. Here $(E_p^2 dN/dE_p)_\mathrm{obs}$ denotes the observed, approximately isotropic, GCR proton energy spectrum from the Alpha Magnetic Spectrometer \citep[AMS;][]{Aguilar2021} for energies $\lesssim 1$ TeV and from the ISS-CREAM Experiment \citep{Choi2022} at higher energies. 
The final gamma-ray flux expression is $\Phi_\gamma (E_\gamma) = 2 \pi R_\odot^2/L^2  \times (R_\odot/R_\mathrm{1AU})^2 \times \Phi^*_\gamma (E_\gamma)$,
where $R_\mathrm{1AU}$ is the distance between the Sun and Earth.

\section{Results}
\label{sec:results}

\subsection{Interacting vs escaping GCRs}
\label{sec:trend}

\begin{figure*}
    \centering
    \includegraphics[width=0.45\textwidth]{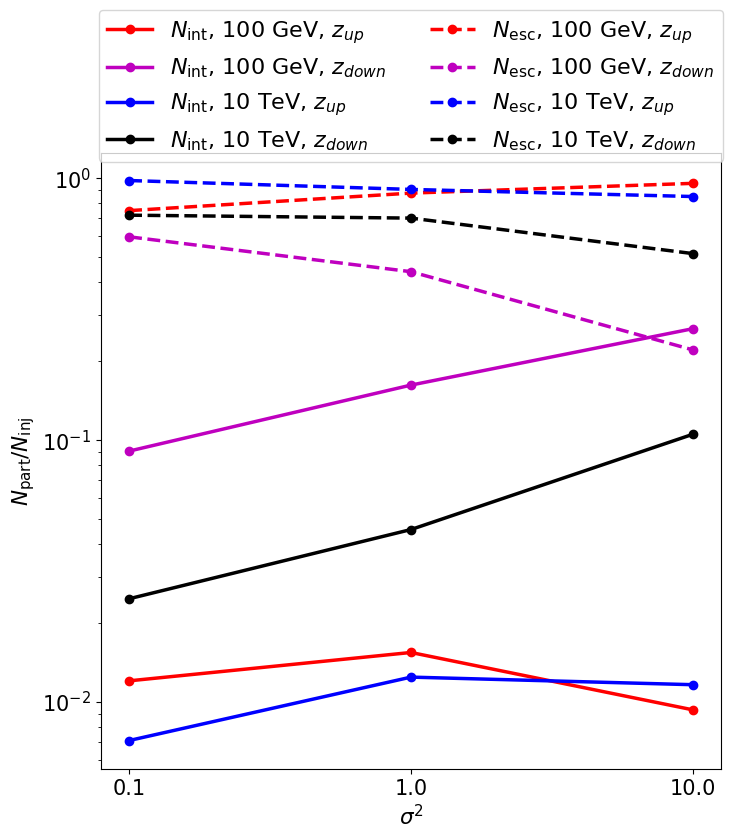}
    \includegraphics[width=0.54\textwidth]{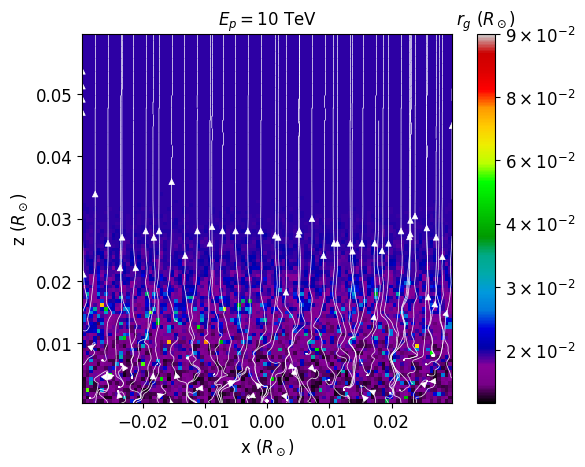}
    
    \caption{\textit{Left panel:} Ratio $N_\mathrm{int}/N_\mathrm{inj}$ ($N_\mathrm{esc}/N_\mathrm{inj}$)  for the interacting (escaping) GCRs marked by the solid (dashed) lines as a function of $\sigma^2$ for injection at $z = z_{up}$ and $z = z_{down}$ for $100$ GeV (red and magenta lines, respectively) and $10$ TeV (blue and black lines, respectively) GCRs.
    \textit{Right panel:} Same as Fig. \ref{fig:laminar.png}, lower left panel, with colorscale showing the gyroradius of $10$ TeV GCRs with magnetic field lines (in white).
    }
    \label{fig:trend}
\end{figure*}

The left panel of Figure \ref{fig:trend} shows $N_\mathrm{int}$ (solid lines) and $N_\mathrm{esc}$ (dashed lines) divided by $N_\mathrm{inj}$ as a function of $\sigma^2$.
The red (blue) and magenta (black) lines show the results for injection at $z = z_{up}$ and $z = z_{down}$, respectively, for $100$ GeV ($10$ TeV) GCRs. 
Due to the periodic boundary conditions in the $xy$-plane, protons can escape only along the $z$-direction (i.e., move outward in the corona or precipitate into the Sun surface).
The panel shows that when particles are injected at $z = z_{\text{down}}$, $N_\mathrm{int}$ increases with $\sigma^2$, as a result of trapping within smaller-scale magnetic structures.
Consequently, $N_\mathrm{esc}$ decreases (see the corresponding dashed lines). 
For protons injected at $z = z_{\text{up}}$, $N_\mathrm{int}$  remains relatively small and approximately constant across all values of $\sigma^2$.
In this scenario, most GCRs escape the computational domain before interacting, as $N_\text{esc}/N_\text{inj} \sim 1$, freely-streaming out from the top of the box due to positive vertical velocities ($v_z > 0$, in half of GCRs at injection) along the laminar magnetic field due to the large $r_g$ (see right panel of Fig. \ref{fig:trend}).


In this scenario, field distortions in the $xy$-plane increase the residence time of GCRs in the dense photo- and chromo-sphere.
This condition enhances trapping and $\gamma$-ray emission for higher $\sigma^2$. In \cite{Puzzoni2024}, a large-scale non-uniform unperturbed field $B_0(\mathbf{x})$ is overlapped to ubiquitous small-scale fluctuations: $\lambda_\parallel$ is finite everywhere and scales inversely with fluctuation power \citep{{Fraschetti2022}}.  
In that magnetic field geometry, a larger cross-field diffusion \citep[][]{Fraschetti2011} enables horizontal migration 
 to the arcades and a smaller $\lambda_\parallel$ favors the GCR-proton collision (higher $N_\mathrm{int}$). 

\subsection{Angular distribution of outgoing $\gamma$-rays}
\label{sec:angular}
\begin{figure*}
     \centering

     \includegraphics[width=0.49\textwidth]{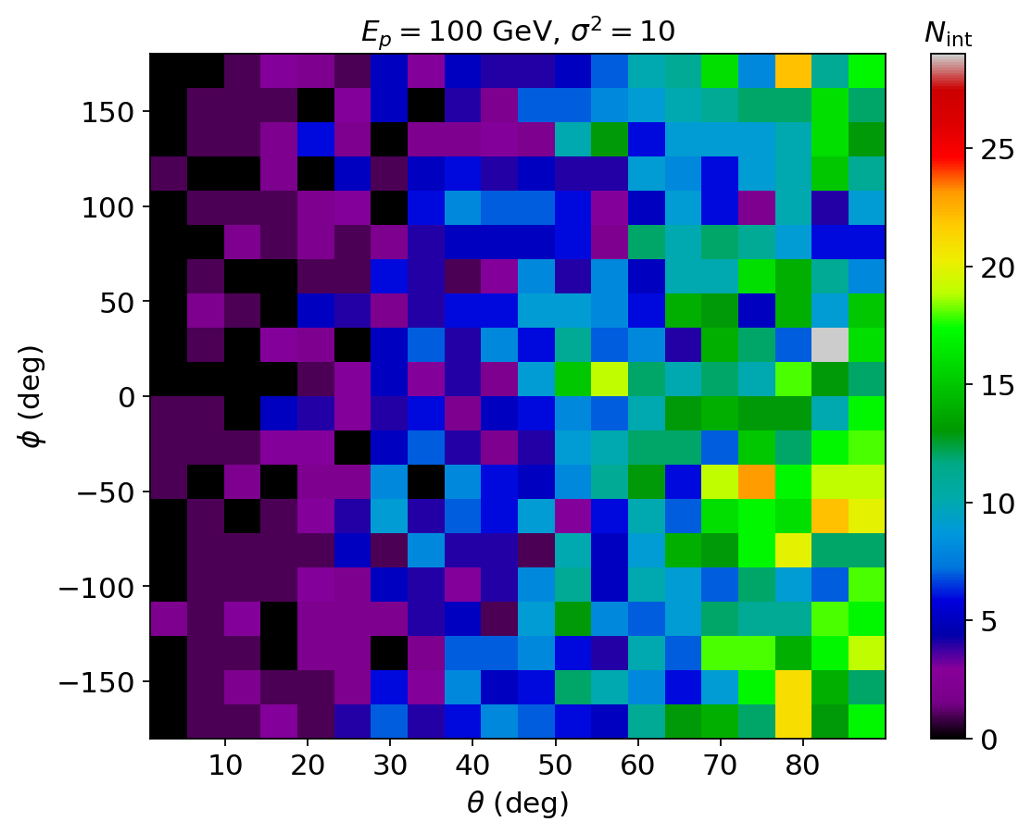} 
      \includegraphics[width=0.49\textwidth]{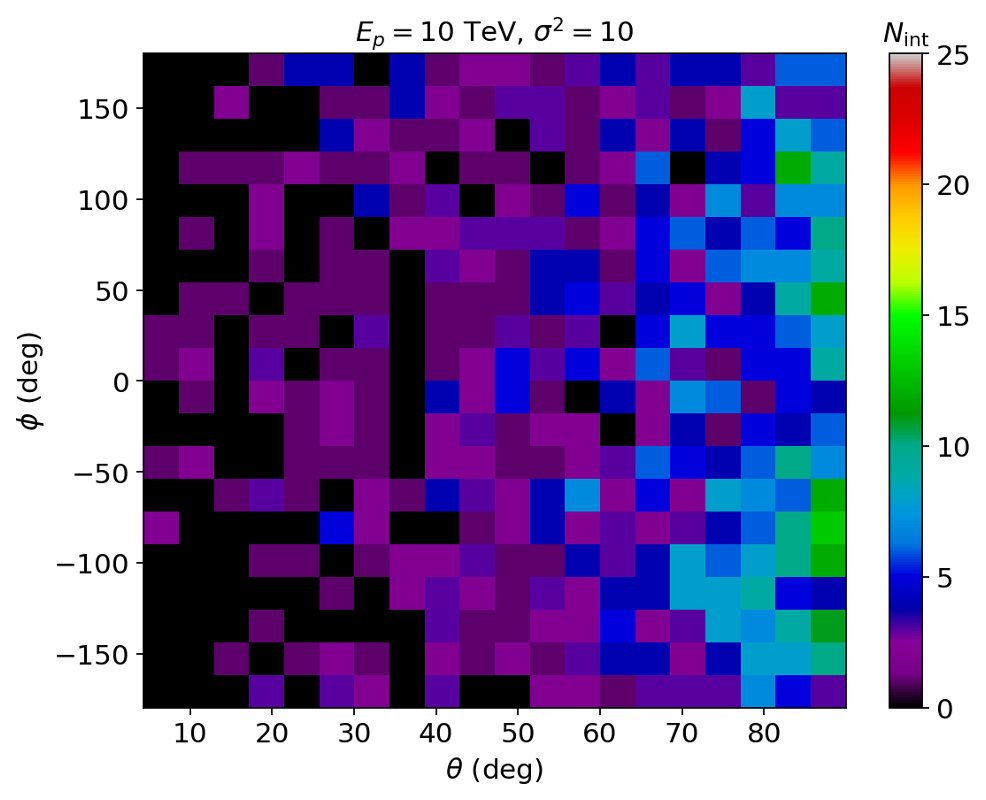}
    
    \caption{\textit{Left panel}: Angular component $\theta$ and $\phi$ of the velocity vector of the GCRs at $t_\mathrm{int}$  for $E_p = 100$ GeV with $\sigma^2 = 10$ and injection at $z = z_\mathrm{down}$. \textit{Right panel:} Same as left panel for $E_p = 10$ TeV. 
    }
    \label{fig:phi}
\end{figure*}

Figure \ref{fig:phi} shows the $0^\circ < \theta < 90^\circ$ and $-180^\circ < \phi < 180^\circ$ angles of the GCR velocity vector at time $t_{int}$ for the $100$ GeV (left panel) and $10$ TeV (right panel) case for $\sigma^2 = 10$ and for injection at $z = z_{down}$.
Due to momentum conservation, the direction of the outgoing GCR is nearly-aligned with the direction of the outgoing photons in the solar plasma frame. 
\cite{LiZhe2024} demonstrated that the angular distribution broadens for $E_p \lesssim 100$ GeV (or $E_\gamma \lesssim 10$ GeV); such an effect is neglected herein as it introduces only a correction smaller by a factor $\sim 4 - 5$ for $E_p < 10$ GeV and even smaller at $E_p > 10$ GeV. 

Figure \ref{fig:phi} also shows, for $E_p = 10$ TeV (right panel), 
that most GCRs interact across the entire $\phi$ range, within the narrow range $ 70^\circ \lesssim \theta \lesssim 90^\circ$ nearly aligned with the tangent plane (i.e., $xy$-plane, see Fig.\ref{fig:domain}), whereas for $E_p = 100$ GeV (left panel) the $\theta$ range broadens down to $\theta \approx 50^\circ$. 
This implies that as GCR energy increases, gamma-rays are emitted close to the plane tangent to the Sun surface, as also found in \cite{Puzzoni2024}. 
A similar anisotropic gamma-ray emission pattern on the solar disk is also suggested by Fermi-LAT observations \citep{Linden2022}. 

\subsection{Gamma-rays flux}
\label{sec:flux}

\begin{figure*}
    \centering
    \includegraphics[width=0.99\textwidth]{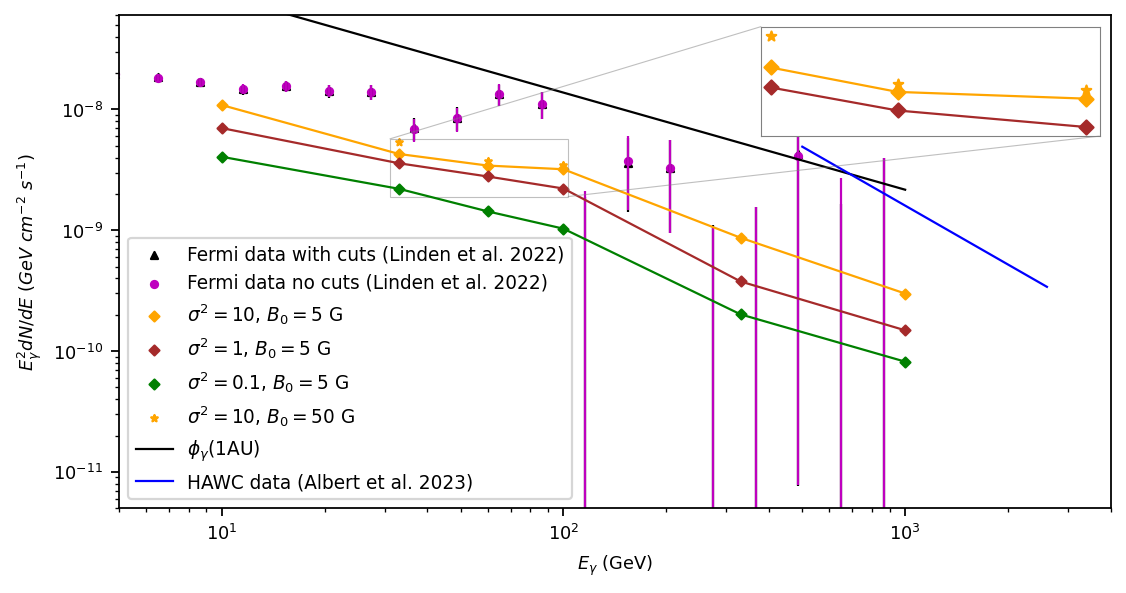}
    
    \caption{Solar disk-average gamma-ray spectrum observed by Fermi-LAT with (purple points) and without (black points) solar flares, from \cite{Linden2022}, by HAWC \citep[blue line, from][]{Albert2023}, and obtained from our simulations with $\sigma^2 = 10$ (orange diamonds), $\sigma^2 = 1$ (in brown), and $\sigma^2 = 0.1$ (in green). The orange stars represent simulations with $\sigma^2 =10$ but with a total magnetic field strength 10 times higher ($B_0 = 50$ G leading to low-photosphere $B_{max} \sim 1$ kG). A detailed comparison is shown in the zoomed-in inset at the top right. The solid black line represents the estimated (see text for further details) gamma-ray flux considering a 100\% efficiency of conversion from GCRs to gamma-rays.}
    \label{fig:flux}
\end{figure*}

Figure \ref{fig:flux} presents the solar disk gamma-ray flux derived from our simulations for $B_0=5$ G and different values of $\sigma^2$: 0.1 (green), 1 (brown), and 10 (orange diamonds, leading to a low-photosphere $B_\mathrm{max} \sim 100$ G). The orange stars represent results obtained with $\sigma^2 = 10$ and $B_0=50$ G, leading to $B_\mathrm{max} \sim 1$ kG.
These results are compared with the Fermi-LAT data \citep{Linden2022} with and without the solar flares (purple and black points, respectively), with the HAWC observations \citep[in blue,][]{Albert2023}, and the estimated gamma-ray flux at 1 AU assuming $100\%$ efficiency in converting protons to gamma-rays (solid black line). The latter is calculated using the expression: $\phi_\gamma (1AU) \approx (E_p^2 dN/dE_p)_\mathrm{obs} (R_\odot/R_\mathrm{1AU})^2(E_\gamma/E_p)$, where $E_\gamma/E_p \sim 1/10$.
Results at energies $E_\gamma < 10$ GeV (corresponding to $E_p < 100$ GeV) are omitted, as the gyromotion of particles in this energy range cannot be resolved given the spatial resolution of $\simeq 418$ km.

The gamma-ray flux increases with $\sigma^2$ across the entire energy range analyzed due to the increase in $N_\mathrm{int}$ (see left panel of Fig. \ref{fig:trend}).
Small-scale magnetic structures in the low photosphere effectively trap GCRs, extending their residence time and consequently increasing the probability of interaction.
This finding aligns with expectations and our earlier work 
\citep{Puzzoni2024}.
A notable difference from the arcade trapping of GCRs analyzed in \cite{Puzzoni2024} is that, at higher energies ($E_\gamma \gtrsim 100$ GeV), the gamma-ray flux from an arcade does not depend on the amplitude of turbulent fluctuations. That result indicates that the gamma-ray flux is due to the migration from open to closed field lines: once a GCR is trapped in the arcade the flux is independent of the turbulence strength, and drops as a result of the drop of the incoming GCRs flux unable to enter the arcade.

With the specific magnetic field geometry in the present paper, we observe a novel trend in the ``dip'' energy range (from $\sim 30$ up to $100$ GeV) as $\sigma^2$ increases, as highlighted in the zoomed-in inset in the top right of Fig. \ref{fig:flux}.
For $\sigma^2 \le 0.1$, the gamma-ray flux decreases in this region, at a rate comparable to the maximum efficiency spectrum (solid black line).
As $\sigma^2$ increases, however, the gamma-ray flux begins to flatten within this range, even for $B_0$ ten times stronger.
As elaborated in the next section, this effect is not numerical, but genuinely physical.

\section{Discussion and conclusions}
\label{sec:summary}
%
%
This paper extends the previous work on the role of closed magnetic arcades in the solar disk gamma-ray emission presented in \cite{Puzzoni2024} by using a new model for the magnetic field near the solar surface which has the property of being open to space at the top, and increasingly horizontal near the surface. In our new model, the GCRs follow the open field lines from the top towards the solar surface where they encounter an irregular field \citep[see][]{Giacalone2021}.

In agreement with \cite{Li2024} and \cite{Puzzoni2024}, GCR protons predominantly interact with solar atmospheric protons in the lower dense solar photosphere \citep[see, e.g.,][]{Morton2023}.
$N_\mathrm{int}$ increases as expected with the field irregularity ($\sigma^2$) if GCRs are injected within the non-vertical field layer. However,
$N_\mathrm{int}$ is approximately independent on $\sigma^2$ for injection further up in the laminar field region. 
In this case, the majority of high-energy GCRs ($E_p  \gtrsim 10$ TeV) escape upward into the lower corona due to their large $r_g$ \citep[as sideway escape is treated via periodic boundary conditions in the $xy$-plane, unlike in][]{Puzzoni2024}.
To clarify this, we injected GCRs at lower altitudes while remaining within the laminar field region (specifically, at $z = 0.025 R_\odot$ instead of $z = 0.05 R_\odot$). 
This adjustment resulted in an increase in $N_\mathrm{int}$, though it had a minimal impact on the intensity of the gamma-ray flux.

It is worthwhile to note that our simulations are consistent with the GCRs pitch-angle isotropy \citep[for ACRs see][]{Rankin2024}: if initially only $v_z <0$ is allowed, the resulting gamma-ray flux would be up to five times higher than our calculated low-energy flux, while the spectral slope would remain approximately unchanged. An isotropic distribution is also assumed in a study focusing on TeV energy range only \citep{Ng2024}.

Gamma-ray emission spans the full range in the $\phi$ angular component of the outgoing GCR at all energies (see Fig.\ref{fig:phi}). At higher energies, a gamma-ray ``trapping cone'', i.e., angular range from the vertical to the Sun surface where proton escape is prohibited, extends down to  $\theta \approx 90^\circ$ (along the $xy$-plane); thus, gamma-ray emission is squeezed to the tangent plane. At lower-energy, the trapping cone is narrower, gamma-rays are emitted at smaller $\theta$ and the solar disk emission appears more isotropic.
This energy-dependent emission pattern on the solar disk is roughly consistent with Fermi-LAT observations \citep[see][]{Linden2022} and numerical results reported in \cite{Puzzoni2024}.

The new synthetic magnetic field model also reflects a realistic solar atmosphere, characterized by nearly-closed magnetic structures extending to altitudes on the order of Mm, as observed by Solar Orbiter \citep{Antolin.etal:23}, with maximum magnetic field strengths ranging from $100$ G up to $1$ kG and profile of the horizontal field both consistent with remote sensing observations \citep[see, e.g.,][]{Danilovic2010, Stenflo2013} and MHD simulations \citep[see, e.g.,][and references therein]{Rempel2014}.
Rather than deriving the field from a specific physical mechanism, we constructed it to mimic a geometry consistent with observations and likely produced by processes such as granular flows, MHD turbulence, or reconnection-driven braiding, which are known to distort field lines in the low solar atmosphere \citep[see, e.g.,][]{Morton2023}. 
The model is consistent with 3D MHD-based reconstructions  \citep[see, e.g.,][]{Gonzales2018, Hudson2020}.

A highly irregular  photospheric field was essential for revealing features of the gamma-ray flux within the dip region.
In \cite{Puzzoni2024}, we argued that the dip/rebrightening of the gamma-ray flux between approximately $30$ and $100$ GeV can be due to an energy cutoff in the migration of the GCRs into magnetic arcades (for a given turbulence strength, only GCRs with large enough $E_p$ can migrate into the arcades and emit $\gamma$-rays).
Here, we show that as the field becomes more irregular, the gamma-ray flux flattens between $30$ and $100$ GeV. It is noteworthy that this flattening saturates at the shape shown in Figure \ref{fig:flux}, rather than forming a local maximum for $\sigma^2 \gtrsim 10$.
This has been demonstrated by running the $\sigma^2 = 50$ case as well as the case with a stronger magnetic field ($B_\mathrm{max} \sim 1$ kG reached with $B_0 = 50$ G).
Notably, this spectral dip was not captured in earlier modeling attempts \citep[see, e.g.,][with the single-structure model]{Gutierrez2022, Li2024}.

We argue that the dip-rebrightening is associated with the condition $r_g (300 {{\rm GeV}})\simeq L_s$ in a $B_0 = 5$ G field (see Fig. \ref{fig:laminar.png}). 
To investigate this, we break such a condition by increasing $L_s$ to $3L_s \sim \Lambda$, with unchanged $\Lambda$ and $\sigma^2$, thereby increasing the length-scale that maintains coherent the horizontal field. Horizontal GCRs motion for a longer time reduces the escape along the vertical direction, leading to an enhanced flux. 
Our findings show that for $\sigma^2 = 0.1$, the slope of the gamma-ray flux in the dip region remains unchanged, while its intensity increases. The flux does not change significantly for higher $\sigma^2$ (or higher $\langle \theta \rangle_\perp$).
GCR-proton interactions (in the dip energy range) have a mean free path $\lambda_{pp} \simeq (\sigma_\mathrm{pp} \,\rho_0/m_p)^{-1} \lesssim L_s$ for photospheric densities $\rho_0 \sim 10^{-7} - 10^{-6}$ g/cm$^3$ (see Sect. \ref{sec:dens_prof}); thus, the flux at the dip is unaffected by a greater $L_s$. 

In addition, we have examined the effect of $\Lambda$ on the gamma-ray flux. Doubling $\Lambda$ for all the values of $\sigma^2$ with an unchanged density vertical profile, does not alter either the intensity nor the slope of the resulting gamma-ray spectrum at $E_\gamma \gtrsim 30$ GeV.
The reason is that, in our synthetic field, modifying the height of deformation of the vertical field $\Lambda$ does not affect the underlying density profile: the denser region where the majority of GCR-proton collisions take place remains unchanged, as well as the gamma-rays production.
Therefore, we conclude that the dip-rebrightening is not a numerical artifact resulting from the chosen computational setup but rather arises from a physical effect associated with the small scale magnetic field structures.

The present model has limitations: we cannot include closed magnetic field lines, which efficiently trap GCRs enhancing the gamma-ray flux, and the time dependence of the latitudinal emission (equatorial vs polar) over the solar cycle; in addition, smaller-scale structures that contribute low-energy flux ($\lesssim 10$ GeV) are not resolved. 
It is important to underline that this model provides a geometric framework, consistent with observed/simulated irregularity, that explains the gamma-ray production; the magnetic field structure is not evolved via a specific MHD model. 
Identifying which physical processes are responsible for the braiding remains an open question for future work. 
We also assume that the field is static over the GCR propagation timescale (tens of seconds), which is much shorter than the characteristic timescale of quiescent  photo- and chromospheric field evolution \citep[e.g., tens of minutes for the active regions loops,][]{Reale:14}. This justifies neglecting temporal variations in the field while focusing on its spatial structure.
With the inclusion of open magnetic field lines in our modeling, a natural progression is to investigate the effect of interchange reconnection, which could be addressed in future work by including a multi-scale field model with both open and closed topologies, informed by high-resolution MHD simulations. Despite these limitations, this work supports the findings of our previous work \citep{Puzzoni2024}, highlighting the critical role of the photospheric magnetic field in shaping the gamma-ray flux from the solar disk.  Our inclusion of irregular field provides a crucial new perspective, helping to address previously unexplained spectral features, such as the dip observed in the $\sim 30-100$ GeV range, which had remained unexplained in earlier models.

\let\internallinenumbers\relax
\begin{acknowledgments}
The authors are grateful to the referee for a very constructive feedback. The authors express gratitude for the insightful feedback from Drs. M. Asgari-Targhi, J. Beacom, S. S. Cerri, O. Cohen, C. Fong, H. S. Hudson, J.-T. Li, T. Linden, K. C. Y. Ng, A. Peter. This work was partially  supported  by NASA under grant 80NSSC22K0040. F. F. was partially supported by NASA under grants 80NSSC21K0119 and 80NSSC21K1766. E. P., J. K., and J. G. were partially supported by NASA under grant 18-DRIVE18$\_$2-0029. J. G. acknowledges partial support from NSF under grant No. 1931252. 
\end{acknowledgments}

%

\vspace{5mm}
\facilities{Fermi, HAWC}


\software{PLUTO code \citep{Mignone2007, Mignone2012}
          }





\bibliography{sample631}{}

\begin{thebibliography}{}
\expandafter\ifx\csname natexlab\endcsname\relax\def\natexlab#1{#1}\fi
\providecommand{\url}[1]{\href{#1}{#1}}
\providecommand{\dodoi}[1]{doi:~\href{http://doi.org/#1}{\nolinkurl{#1}}}
\providecommand{\doeprint}[1]{\href{http://ascl.net/#1}{\nolinkurl{http://ascl.net/#1}}}
\providecommand{\doarXiv}[1]{\href{https://arxiv.org/abs/#1}{\nolinkurl{https://arxiv.org/abs/#1}}}

\bibitem[{{Abdo} {et~al.}(2011){Abdo}, {Ackermann}, {Ajello}, {Baldini}, {Ballet}, {Barbiellini}, {Bastieri}, {Bechtol}, {Bellazzini}, {Berenji}, {Blandford}, {Bloom}, {Bonamente}, {Borgland}, {Bouvier}, {Bregeon}, {Brez}, {Brigida}, {Bruel}, {Buehler}, {Buson}, {Caliandro}, {Cameron}, {Cannon}, {Caraveo}, {Carrigan}, {Casandjian}, {Cavazzuti}, {Cecchi}, {{\c{C}}elik}, {Charles}, {Chekhtman}, {Chiang}, {Ciprini}, {Claus}, {Cohen-Tanugi}, {Conrad}, {Cutini}, {de Angelis}, {de Palma}, {Dermer}, {Silva}, {Drell}, {Dubois}, {Dumora}, {Escande}, {Favuzzi}, {Fegan}, {Finke}, {Focke}, {Fortin}, {Frailis}, {Fuhrmann}, {Fukazawa}, {Fukuyama}, {Funk}, {Fusco}, {Gargano}, {Gasparrini}, {Gehrels}, {Georganopoulos}, {Germani}, {Giebels}, {Giglietto}, {Giommi}, {Giordano}, {Giroletti}, {Glanzman}, {Godfrey}, {Grenier}, {Guiriec}, {Hadasch}, {Hayashida}, {Hays}, {Horan}, {Hughes}, {J{\'o}hannesson}, {Johnson}, {Johnson}, {Kadler}, {Kamae}, {Katagiri}, {Kataoka}, {Kn{\"o}dlseder}, {Kuss}, {Lande}, {Latronico}, {Lee},
  {Longo}, {Loparco}, {Lott}, {Lovellette}, {Lubrano}, {Madejski}, {Makeev}, {Max-Moerbeck}, {Mazziotta}, {McEnery}, {Mehault}, {Michelson}, {Mitthumsiri}, {Mizuno}, {Monte}, {Monzani}, {Morselli}, {Moskalenko}, {Murgia}, {Nakamori}, {Naumann-Godo}, {Nishino}, {Nolan}, {Norris}, {Nuss}, {Ohsugi}, {Okumura}, {Omodei}, {Orlando}, {Ormes}, {Ozaki}, {Paneque}, {Panetta}, {Parent}, {Pavlidou}, {Pearson}, {Pelassa}, {Pepe}, {Pesce-Rollins}, {Pierbattista}, {Piron}, {Porter}, {Rain{\`o}}, {Rando}, {Razzano}, {Readhead}, {Reimer}, {Reimer}, {Reyes}, {Richards}, {Ritz}, {Roth}, {Sadrozinski}, {Sanchez}, {Sander}, {Sgr{\`o}}, {Siskind}, {Smith}, {Spandre}, {Spinelli}, {Stawarz}, {Stevenson}, {Strickman}, {Suson}, {Takahashi}, {Takahashi}, {Tanaka}, {Thayer}, {Thayer}, {Thompson}, {Tibaldo}, {Torres}, {Tosti}, {Tramacere}, {Troja}, {Usher}, {Vandenbroucke}, {Vasileiou}, {Vianello}, {Vilchez}, {Vitale}, {Waite}, {Wang}, {Wehrle}, {Winer}, {Wood}, {Yang}, {Yatsu}, {Ylinen}, {Zensus}, {Ziegler}, {Fermi LAT Collaboration},
  {Aleksi{\'c}}, {Antonelli}, {Antoranz}, {Backes}, {Barrio}, {Becerra Gonz{\'a}lez}, {Bednarek}, {Berdyugin}, {Berger}, {Bernardini}, {Biland}, {Blanch}, {Bock}, {Boller}, {Bonnoli}, {Bordas}, {Borla Tridon}, {Bosch-Ramon}, {Bose}, {Braun}, {Bretz}, {Camara}, {Carmona}, {Carosi}, {Colin}, {Colombo}, {Contreras}, {Cortina}, {Covino}, {Dazzi}, {de Angelis}, {De Cea del Pozo}, {Delgado Mendez}, {De Lotto}, {De Maria}, {De Sabata}, {Diago Ortega}, {Doert}, {Dom{\'\i}nguez}, {Dominis Prester}, {Dorner}, {Doro}, {Elsaesser}, {Ferenc}, {Fonseca}, {Font}, {Garc{\'\i}a L{\'o}pez}, {Garczarczyk}, {Gaug}, {Giavitto}, {Godinovi}, {Hadasch}, {Herrero}, {Hildebrand}, {H{\"o}hne-M{\"o}nch}, {Hose}, {Hrupec}, {Jogler}, {Klepser}, {Kr{\"a}henb{\"u}hl}, {Kranich}, {Krause}, {La Barbera}, {Leonardo}, {Lindfors}, {Lombardi}, {L{\'o}pez}, {Lorenz}, {Majumdar}, {Makariev}, {Maneva}, {Mankuzhiyil}, {Mannheim}, {Maraschi}, {Mariotti}, {Mart{\'\i}nez}, {Mazin}, {Meucci}, {Miranda}, {Mirzoyan}, {Miyamoto}, {Mold{\'o}n}, {Moralejo},
  {Nieto}, {Nilsson}, {Orito}, {Oya}, {Paoletti}, {Paredes}, {Partini}, {Pasanen}, {Pauss}, {Pegna}, {Perez-Torres}, {Persic}, {Peruzzo}, {Pochon}, {Prada}, {Prada Moroni}, {Prandini}, {Puchades}, {Puljak}, {Reichardt}, {Rhode}, {Rib{\'o}}, {Rico}, {Rissi}, {R{\"u}gamer}, {Saggion}, {Saito}, {Saito}, {Salvati}, {S{\'a}nchez-Conde}, {Satalecka}, {Scalzotto}, {Scapin}, {Schultz}, {Schweizer}, {Shayduk}, {Shore}, {Sierpowska-Bartosik}, {Sillanp{\"a}{\"a}}, {Sitarek}, {Sobczynska}, {Spanier}, {Spiro}, {Stamerra}, {Steinke}, {Storz}, {Strah}, {Struebig}, {Suric}, {Takalo}, {Tavecchio}, {Temnikov}, {Terzi{\'c}}, {Tescaro}, {Teshima}, {Vankov}, {Wagner}, {Weitzel}, {Zabalza}, {Zandanel}, {Zanin}, {MAGIC Collaboration}, {Villata}, {Raiteri}, {Aller}, {Aller}, {Chen}, {Jordan}, {Koptelova}, {Kurtanidze}, {L{\"a}hteenm{\"a}ki}, {McBreen}, {Larionov}, {Lin}, {Nikolashvili}, {Reinthal}, {Angelakis}, {Capalbi}, {Carrami{\~n}ana}, {Carrasco}, {Cassaro}, {Cesarini}, {Falcone}, {Gurwell}, {Hovatta}, {Kovalev}, {Kovalev},
  {Krichbaum}, {Krimm}, {Lister}, {Moody}, {Maccaferri}, {Mori}, {Nestoras}, {Orlati}, {Pace}, {Pagani}, {Pearson}, {Perri}, {Piner}, {Ros}, {Sadun}, {Sakamoto}, {Tammi}, \& {Zook}}]{Abdo2011}
{Abdo}, A.~A., {Ackermann}, M., {Ajello}, M., {et~al.} 2011, \apj, 736, 131, \dodoi{10.1088/0004-637X/736/2/131}

\bibitem[{{Abeysekara} {et~al.}(2013){Abeysekara}, {Alfaro}, {Alvarez}, {{\'A}lvarez}, {Arceo}, {Arteaga-Vel{\'a}zquez}, {Ayala Solares}, {Barber}, {Baughman}, {Bautista-Elivar}, {Belmont}, {BenZvi}, {Berley}, {Bonilla Rosales}, {Braun}, {Caballero-Lopez}, {Carrami{\~n}ana}, {Castillo}, {Cotti}, {Cotzomi}, {de la Fuente}, {De Le{\'o}n}, {DeYoung}, {Diaz Hernandez}, {Diaz-Velez}, {Dingus}, {DuVernois}, {Ellsworth}, {Fernandez}, {Fiorino}, {Fraija}, {Galindo}, {Garcia-Luna}, {Garcia-Torales}, {Garfias}, {Gonz{\'a}lez}, {Gonz{\'a}lez}, {Goodman}, {Grabski}, {Gussert}, {Hampel-Arias}, {Hui}, {H{\"u}ntemeyer}, {Imran}, {Iriarte}, {Karn}, {Kieda}, {Kunde}, {Lara}, {Lauer}, {Lee}, {Lennarz}, {Le{\'o}n Vargas}, {Linares}, {Linnemann}, {Longo}, {Luna-Garc{\'\i}a}, {Marinelli}, {Martinez}, {Mart{\'\i}nez-Castro}, {Matthews}, {Miranda-Romagnoli}, {Moreno}, {Mostaf{\'a}}, {Nava}, {Nellen}, {Newbold}, {Noriega-Papaqui}, {Oceguera-Becerra}, {Patricelli}, {Pelayo}, {P{\'e}rez-P{\'e}rez}, {Pretz}, {Rivi{\`e}re}, {Ryan},
  {Rosa-Gonz{\'a}lez}, {Salazar}, {Salesa}, {Sandoval}, {Santos}, {Schneider}, {Silich}, {Sinnis}, {Smith}, {Sparks}, {Springer}, {Taboada}, {Toale}, {Tollefson}, {Torres}, {Ukwatta}, {Villase{\~n}or}, {Weisgarber}, {Westerhoff}, {Wisher}, {Wood}, {Yodh}, {Younk}, {Zaborov}, {Zepeda}, \& {Zhou}}]{Abeysekara2013}
{Abeysekara}, A.~U., {Alfaro}, R., {Alvarez}, C., {et~al.} 2013, Astroparticle Physics, 50, 26, \dodoi{10.1016/j.astropartphys.2013.08.002}

\bibitem[{{Acharyya} {et~al.}(2025){Acharyya}, {Adelfio}, {Ajello}, {Baldini}, {Bartolini}, {Bastieri}, {Becerra Gonzalez}, {Bellazzini}, {Berenji}, {Bissaldi}, {Blandford}, {Bonino}, {Bottacini}, {Buson}, {Cameron}, {Caraveo}, {Casaburo}, {Casini}, {Cavazzuti}, {Cerasole}, {Ciprini}, {Cozzolongo}, {Cristarella Orestano}, {Cuoco}, {Cutini}, {D'Ammando}, {Depalo}, {Digel}, {Di Lalla}, {Di Venere}, {Dom{\'\i}nguez}, {Fiori}, {Fukazawa}, {Fusco}, {Gargano}, {Gasbarra}, {Gasparrini}, {Germani}, {Giacchino}, {Giglietto}, {Giordano}, {Giroletti}, {Guiriec}, {Gupta}, {Hashizume}, {Hays}, {Hewitt}, {Holzmann Airasca}, {Horan}, {Hou}, {Kayanoki}, {Kuss}, {Larsson}, {Laviron}, {Li}, {Liodakis}, {Longo}, {Loparco}, {L{\'o}pez P{\'e}rez}, {Lovellette}, {Lubrano}, {Maldera}, {Manfreda}, {Mart{\'\i}-Devesa}, {Martinelli}, {Mazziotta}, {McEnery}, {Mereu}, {Michailidis}, {Michelson}, {Mirabal}, {Mizuno}, {Monti-Guarnieri}, {Monzani}, {Morselli}, {Moskalenko}, {Omodei}, {Orlando}, {Ormes}, {Paneque}, {Persic},
  {Pesce-Rollins}, {Petrosian}, {Pillera}, {Principe}, {Rain{\`o}}, {Rando}, {Rani}, {Razzano}, {Reimer}, {Reimer}, {S{\'a}nchez-Conde}, {Saz Parkinson}, {Serini}, {Sgr{\`o}}, {Siskind}, {Spinelli}, {Tak}, {Tibaldo}, {Torres}, {Valverde}, {Wadiasingh}, \& {Zhang}}]{Acharyya2025}
{Acharyya}, A., {Adelfio}, A., {Ajello}, M., {et~al.} 2025, arXiv e-prints, arXiv:2505.06348, \dodoi{10.48550/arXiv.2505.06348}

\bibitem[{{Aguilar} {et~al.}(2021){Aguilar}, {Ali Cavasonza}, {Ambrosi}, {Arruda}, {Attig}, {Barao}, {Barrin}, {Bartoloni}, {Ba{\c{s}}e{\u{g}}mez-du Pree}, {Bates}, {Battiston}, {Behlmann}, {Beischer}, {Berdugo}, {Bertucci}, {Bindi}, {de Boer}, {Bollweg}, {Borgia}, {Boschini}, {Bourquin}, {Bueno}, {Burger}, {Burger}, {Burmeister}, {Cai}, {Capell}, {Casaus}, {Castellini}, {Cervelli}, {Chang}, {Chen}, {Chen}, {Chen}, {Cheng}, {Chou}, {Chouridou}, {Choutko}, {Chung}, {Clark}, {Coignet}, {Consolandi}, {Contin}, {Corti}, {Cui}, {Dadzie}, {Dai}, {Delgado}, {Della Torre}, {Demirk{\"o}z}, {Derome}, {Di Falco}, {Di Felice}, {D{\'\i}az}, {Dimiccoli}, {von Doetinchem}, {Dong}, {Donnini}, {Duranti}, {Egorov}, {Eline}, {Feng}, {Fiandrini}, {Fisher}, {Formato}, {Freeman}, {Galaktionov}, {G{\'a}mez}, {Garc{\'\i}a-L{\'o}pez}, {Gargiulo}, {Gast}, {Gebauer}, {Gervasi}, {Giovacchini}, {G{\'o}mez-Coral}, {Gong}, {Goy}, {Grabski}, {Grandi}, {Graziani}, {Guo}, {Haino}, {Han}, {Hashmani}, {He}, {Heber}, {Hsieh}, {Hu}, {Huang},
  {Hungerford}, {Incagli}, {Jang}, {Jia}, {Jinchi}, {Kanishev}, {Khiali}, {Kim}, {Kirn}, {Konyushikhin}, {Kounina}, {Kounine}, {Koutsenko}, {Kuhlman}, {Kulemzin}, {La Vacca}, {Laudi}, {Laurenti}, {Lazzizzera}, {Lebedev}, {Lee}, {Lee}, {Leluc}, {Li}, {Li}, {Li}, {Li}, {Li}, {Li}, {Light}, {Lin}, {Lippert}, {Liu}, {Lu}, {Lu}, {Luebelsmeyer}, {Luo}, {Lyu}, {Machate}, {Ma{\~n}{\'a}}, {Mar{\'\i}n}, {Marquardt}, {Martin}, {Mart{\'\i}nez}, {Masi}, {Maurin}, {Menchaca-Rocha}, {Meng}, {Mo}, {Molero}, {Mott}, {Mussolin}, {Ni}, {Nikonov}, {Nozzoli}, {Oliva}, {Orcinha}, {Palermo}, {Palmonari}, {Paniccia}, {Pashnin}, {Pauluzzi}, {Pensotti}, {Phan}, {Plyaskin}, {Pohl}, {Porter}, {Qi}, {Qin}, {Qu}, {Quadrani}, {Rancoita}, {Rapin}, {Reina Conde}, {Rosier-Lees}, {Rozhkov}, {Rozza}, {Sagdeev}, {Schael}, {Schmidt}, {Schulz von Dratzig}, {Schwering}, {Seo}, {Shan}, {Shi}, {Siedenburg}, {Solano}, {Song}, {Sonnabend}, {Sun}, {Sun}, {Tacconi}, {Tang}, {Tang}, {Tian}, {Ting}, {Ting}, {Tomassetti}, {Torsti}, {T{\"u}ys{\"u}z},
  {Urban}, {Usoskin}, {Vagelli}, {Vainio}, {Valente}, {Valtonen}, {V{\'a}zquez Acosta}, {Vecchi}, {Velasco}, {Vialle}, {Wang}, {Wang}, {Wang}, {Wang}, {Wang}, {Wang}, {Wei}, {Weng}, {Wu}, {Xiong}, {Xu}, {Yan}, {Yang}, {Yi}, {Yu}, {Yu}, {Zannoni}, {Zhang}, {Zhang}, {Zhang}, {Zhang}, {Zhang}, {Zhao}, {Zheng}, {Zhuang}, {Zhukov}, {Zichichi}, {Zimmermann}, {Zuccon}, \& {AMS Collaboration}}]{Aguilar2021}
{Aguilar}, M., {Ali Cavasonza}, L., {Ambrosi}, G., {et~al.} 2021, \physrep, 894, 1, \dodoi{10.1016/j.physrep.2020.09.003}

\bibitem[{{Albert} {et~al.}(2023){Albert}, {Alfaro}, {Alvarez}, {Arteaga-Vel{\'a}zquez}, {Avila Rojas}, {Ayala Solares}, {Babu}, {Belmont-Moreno}, {Brisbois}, {Caballero-Mora}, {Capistr{\'a}n}, {Carrami{\~n}ana}, {Casanova}, {Chaparro-Amaro}, {Cotti}, {Cotzomi}, {Couti{\~n}o de Le{\'o}n}, {De la Fuente}, {Diaz Hernandez}, {Dingus}, {DuVernois}, {Durocher}, {D{\'\i}az-V{\'e}lez}, {Ellsworth}, {Engel}, {Espinoza}, {Fan}, {Fang}, {Fern{\'a}ndez Alonso}, {Fleischhack}, {Fraija}, {Garc{\'\i}a-Gonz{\'a}lez}, {Garfias}, {Gonz{\'a}lez}, {Goodman}, {Harding}, {Hernandez}, {Hinton}, {Huang}, {Hueyotl-Zahuantitla}, {H{\"u}ntemeyer}, {Iriarte}, {Joshi}, {Kaufmann}, {Lee}, {Linnemann}, {Longinotti}, {Luis-Raya}, {Malone}, {Martinez}, {Mart{\'\i}nez-Castro}, {Matthews}, {Miranda-Romagnoli}, {Morales-Soto}, {Moreno}, {Mostaf{\'a}}, {Nayerhoda}, {Nellen}, {Nisa}, {Noriega-Papaqui}, {Olivera-Nieto}, {Omodei}, {P{\'e}rez Araujo}, {P{\'e}rez-P{\'e}rez}, {Rho}, {Rosa-Gonz{\'a}lez}, {Ruiz-Velasco}, {Salazar}, {Salazar-Gallegos},
  {Sandoval}, {Schneider}, {Serna-Franco}, {Smith}, {Son}, {Springer}, {Tibolla}, {Tollefson}, {Torres}, {Torres-Escobedo}, {Turner}, {Ure{\~n}a-Mena}, {Varela}, {Villase{\~n}or}, {Wang}, {Watson}, {Willox}, {Yun-C{\'a}rcamo}, {Zhou}, {de Le{\'o}n}, {Beacom}, {Linden}, {Ng}, {Peter}, {Zhou}, \& {HAWC Collaboration}}]{Albert2023}
{Albert}, A., {Alfaro}, R., {Alvarez}, C., {et~al.} 2023, \prl, 131, 051201, \dodoi{10.1103/PhysRevLett.131.051201}

\bibitem[{{Alfaro} {et~al.}(2024){Alfaro}, {Alvarez}, {Arteaga-Vel{\'a}zquez}, {Arunbabu}, {Avila Rojas}, {Babu}, {Belmont-Moreno}, {Caballero-Mora}, {Capistr{\'a}n}, {Carrami{\~n}ana}, {Casanova}, {Col{\'\i}n-Farias}, {Cotti}, {Cotzomi}, {Couti{\~n}o de Le{\'o}n}, {De la Fuente}, {de Le{\'o}n}, {Depaoli}, {Diaz Hernandez}, {D{\'\i}az-V{\'e}lez}, {Durocher}, {DuVernois}, {Engel}, {Espinoza}, {Fan}, {Fraija}, {Garc{\'\i}a-Gonz{\'a}lez}, {Garfias}, {Gonzalez Mu{\~n}oz}, {Gonz{\'a}lez}, {Goodman}, {Harding}, {Huang}, {Hueyotl-Zahuantitla}, {Iriarte}, {Kaufmann}, {Lara}, {Lee}, {Le{\'o}n Vargas}, {Longinotti}, {Luis-Raya}, {Malone}, {Martinez}, {Mart{\'\i}nez-Castro}, {Matthews}, {Miranda-Romagnoli}, {Moreno}, {Mostaf{\'a}}, {Nayerhoda}, {Nellen}, {Niembro}, {Noriega-Papaqui}, {Omodei}, {P{\'e}rez-P{\'e}rez}, {Rho}, {Rosa-Gonz{\'a}lez}, {Ruiz-Velasco}, {Ryan}, {Salazar}, {Salazar-Gallegos}, {Sandoval}, {Serna-Franco}, {Smith}, {Son}, {Springer}, {Tibolla}, {Tollefson}, {Torres}, {Turner}, {Ure{\~n}a-Mena},
  {Varela}, {Villase{\~n}or}, {Wang}, {Watson}, {Willox}, {Yun-C{\'a}rcamo}, \& {Zhou}}]{Alfaro2024}
{Alfaro}, R., {Alvarez}, C., {Arteaga-Vel{\'a}zquez}, J.~C., {et~al.} 2024, \apj, 966, 67, \dodoi{10.3847/1538-4357/ad3208}

\bibitem[{{Amenomori} {et~al.}(2018){Amenomori}, {Bi}, {Chen}, {Chen}, {Chen}, {Cui}, {Danzengluobu}, {Feng}, {Feng}, {Feng}, {Gou}, {Guo}, {He}, {He}, {Hibino}, {Hotta}, {Hu}, {Hu}, {Huang}, {Jia}, {Jiang}, {Kajino}, {Kasahara}, {Katayose}, {Kato}, {Kawata}, {Kozai}, {Labaciren}, {Li}, {Li}, {Li}, {Liu}, {Liu}, {Liu}, {Lu}, {Meng}, {Miyazaki}, {Mizutani}, {Munakata}, {Nakajima}, {Nakamura}, {Nanjo}, {Nishizawa}, {Niwa}, {Ohnishi}, {Ohta}, {Ozawa}, {Qian}, {Qu}, {Saito}, {Saito}, {Sakata}, {Sako}, {Shao}, {Shibata}, {Shiomi}, {Shirai}, {Sugimoto}, {Takita}, {Tan}, {Tateyama}, {Torii}, {Tsuchiya}, {Udo}, {Wang}, {Wu}, {Xue}, {Yamamoto}, {Yamauchi}, {Yang}, {Yuan}, {Yuda}, {Zhai}, {Zhang}, {Zhang}, {Zhang}, {Zhang}, {Zhang}, {Zhang}, {Zhaxisangzhu}, \& {Tibet AS {\ensuremath{\gamma}} Collaboration}}]{Amenomori2018}
{Amenomori}, M., {Bi}, X.~J., {Chen}, D., {et~al.} 2018, \prl, 120, 031101, \dodoi{10.1103/PhysRevLett.120.031101}

\bibitem[{{Antolin} {et~al.}(2023){Antolin}, {Dolliou}, {Auch{\`e}re}, {Chitta}, {Parenti}, {Berghmans}, {Aznar Cuadrado}, {Barczynski}, {Gissot}, {Harra}, {Huang}, {Janvier}, {Kraaikamp}, {Long}, {Mandal}, {Peter}, {Rodriguez}, {Sch{\"u}hle}, {Smith}, {Solanki}, {Stegen}, {Teriaca}, {Verbeeck}, {West}, {Zhukov}, {Appourchaux}, {Aulanier}, {Buchlin}, {Delmotte}, {Gilles}, {Haberreiter}, {Halain}, {Heerlein}, {Hochedez}, {Gyo}, {Poedts}, \& {Rochus}}]{Antolin.etal:23}
{Antolin}, P., {Dolliou}, A., {Auch{\`e}re}, F., {et~al.} 2023, \aap, 676, A112, \dodoi{10.1051/0004-6361/202346016}

\bibitem[{{Arsioli} \& {Orlando}(2024)}]{Arsioli2024}
{Arsioli}, B., \& {Orlando}, E. 2024, \apj, 962, 52, \dodoi{10.3847/1538-4357/ad1bd2}

\bibitem[{{Athay}(1976)}]{Athay:76}
{Athay}, R.~G. 1976, {The solar chromosphere and corona: Quiet sun}, Vol.~53, \dodoi{10.1007/978-94-010-1715-2}

\bibitem[{{Atwood} {et~al.}(2009){Atwood}, {Abdo}, {Ackermann}, {Althouse}, {Anderson}, {Axelsson}, {Baldini}, {Ballet}, {Band}, {Barbiellini}, {Bartelt}, {Bastieri}, {Baughman}, {Bechtol}, {B{\'e}d{\'e}r{\`e}de}, {Bellardi}, {Bellazzini}, {Berenji}, {Bignami}, {Bisello}, {Bissaldi}, {Blandford}, {Bloom}, {Bogart}, {Bonamente}, {Bonnell}, {Borgland}, {Bouvier}, {Bregeon}, {Brez}, {Brigida}, {Bruel}, {Burnett}, {Busetto}, {Caliandro}, {Cameron}, {Caraveo}, {Carius}, {Carlson}, {Casandjian}, {Cavazzuti}, {Ceccanti}, {Cecchi}, {Charles}, {Chekhtman}, {Cheung}, {Chiang}, {Chipaux}, {Cillis}, {Ciprini}, {Claus}, {Cohen-Tanugi}, {Condamoor}, {Conrad}, {Corbet}, {Corucci}, {Costamante}, {Cutini}, {Davis}, {Decotigny}, {DeKlotz}, {Dermer}, {de Angelis}, {Digel}, {do Couto e Silva}, {Drell}, {Dubois}, {Dumora}, {Edmonds}, {Fabiani}, {Farnier}, {Favuzzi}, {Flath}, {Fleury}, {Focke}, {Funk}, {Fusco}, {Gargano}, {Gasparrini}, {Gehrels}, {Gentit}, {Germani}, {Giebels}, {Giglietto}, {Giommi}, {Giordano}, {Glanzman},
  {Godfrey}, {Grenier}, {Grondin}, {Grove}, {Guillemot}, {Guiriec}, {Haller}, {Harding}, {Hart}, {Hays}, {Healey}, {Hirayama}, {Hjalmarsdotter}, {Horn}, {Hughes}, {J{\'o}hannesson}, {Johansson}, {Johnson}, {Johnson}, {Johnson}, {Johnson}, {Kamae}, {Katagiri}, {Kataoka}, {Kavelaars}, {Kawai}, {Kelly}, {Kerr}, {Klamra}, {Kn{\"o}dlseder}, {Kocian}, {Komin}, {Kuehn}, {Kuss}, {Landriu}, {Latronico}, {Lee}, {Lee}, {Lemoine-Goumard}, {Lionetto}, {Longo}, {Loparco}, {Lott}, {Lovellette}, {Lubrano}, {Madejski}, {Makeev}, {Marangelli}, {Massai}, {Mazziotta}, {McEnery}, {Menon}, {Meurer}, {Michelson}, {Minuti}, {Mirizzi}, {Mitthumsiri}, {Mizuno}, {Moiseev}, {Monte}, {Monzani}, {Moretti}, {Morselli}, {Moskalenko}, {Murgia}, {Nakamori}, {Nishino}, {Nolan}, {Norris}, {Nuss}, {Ohno}, {Ohsugi}, {Omodei}, {Orlando}, {Ormes}, {Paccagnella}, {Paneque}, {Panetta}, {Parent}, {Pearce}, {Pepe}, {Perazzo}, {Pesce-Rollins}, {Picozza}, {Pieri}, {Pinchera}, {Piron}, {Porter}, {Poupard}, {Rain{\`o}}, {Rando}, {Rapposelli}, {Razzano},
  {Reimer}, {Reimer}, {Reposeur}, {Reyes}, {Ritz}, {Rochester}, {Rodriguez}, {Romani}, {Roth}, {Russell}, {Ryde}, {Sabatini}, {Sadrozinski}, {Sanchez}, {Sander}, {Sapozhnikov}, {Parkinson}, {Scargle}, {Schalk}, {Scolieri}, {Sgr{\`o}}, {Share}, {Shaw}, {Shimokawabe}, {Shrader}, {Sierpowska-Bartosik}, {Siskind}, {Smith}, {Smith}, {Spandre}, {Spinelli}, {Starck}, {Stephens}, {Strickman}, {Strong}, {Suson}, {Tajima}, {Takahashi}, {Takahashi}, {Tanaka}, {Tenze}, {Tether}, {Thayer}, {Thayer}, {Thompson}, {Tibaldo}, {Tibolla}, {Torres}, {Tosti}, {Tramacere}, {Turri}, {Usher}, {Vilchez}, {Vitale}, {Wang}, {Watters}, {Winer}, {Wood}, {Ylinen}, \& {Ziegler}}]{Atwood2009}
{Atwood}, W.~B., {Abdo}, A.~A., {Ackermann}, M., {et~al.} 2009, \apj, 697, 1071, \dodoi{10.1088/0004-637X/697/2/1071}

\bibitem[{{Becker Tjus} {et~al.}(2020){Becker Tjus}, {Desiati}, {D{\"o}pper}, {Fichtner}, {Kleimann}, {Kroll}, \& {Tenholt}}]{Becker2020}
{Becker Tjus}, J., {Desiati}, P., {D{\"o}pper}, N., {et~al.} 2020, \aap, 633, A83, \dodoi{10.1051/0004-6361/201936306}

\bibitem[{{Birdsall} \& {Langdon}(1991)}]{Birdsall1991}
{Birdsall}, C.~K., \& {Langdon}, A.~B. 1991, {Plasma Physics via Computer Simulation}

\bibitem[{{Cannady}(2022)}]{Cannady2022}
{Cannady}, N.~W. 2022, in 37th International Cosmic Ray Conference, 604, \dodoi{10.22323/1.395.0604}

\bibitem[{{Choi} {et~al.}(2022){Choi}, {Seo}, {Aggarwal}, {Amare}, {Angelaszek}, {Bowman}, {Chen}, {Copley}, {Derome}, {Eraud}, {Falana}, {Gerrety}, {Han}, {Huh}, {Haque}, {Hwang}, {Hyun}, {Jeon}, {Jeon}, {Jeong}, {Kang}, {Kim}, {Kim}, {Kim}, {Lee}, {Lee}, {Lee}, {Lu}, {Lundquist}, {Lutz}, {Menchaca-Rocha}, {Ofoha}, {Park}, {Park}, {Park}, {Picot-Clemente}, {Scrandis}, {Smith}, {Takeishi}, {Vedenkin}, {Walpole}, {Weinmann}, {Wu}, {Wu}, {Yin}, {Yoon}, \& {Zhang}}]{Choi2022}
{Choi}, G.~H., {Seo}, E.~S., {Aggarwal}, S., {et~al.} 2022, \apj, 940, 107, \dodoi{10.3847/1538-4357/ac9d2c}

\bibitem[{{Danilovic} {et~al.}(2010){Danilovic}, {Sch{\"u}ssler}, \& {Solanki}}]{Danilovic2010}
{Danilovic}, S., {Sch{\"u}ssler}, M., \& {Solanki}, S.~K. 2010, \aap, 513, A1, \dodoi{10.1051/0004-6361/200913379}

\bibitem[{{del Pino Alem{\'a}n} {et~al.}(2018){del Pino Alem{\'a}n}, {Trujillo Bueno}, {{\v{S}}t{\v{e}}p{\'a}n}, \& {Shchukina}}]{delPino2018}
{del Pino Alem{\'a}n}, T., {Trujillo Bueno}, J., {{\v{S}}t{\v{e}}p{\'a}n}, J., \& {Shchukina}, N. 2018, \apj, 863, 164, \dodoi{10.3847/1538-4357/aaceab}

\bibitem[{{D{\"o}rner} {et~al.}(2025){D{\"o}rner}, {Morejon}, {Kampert}, \& {Becker Tjus}}]{Dorner2025}
{D{\"o}rner}, J., {Morejon}, L., {Kampert}, K.-H., \& {Becker Tjus}, J. 2025, \jcap, 2025, 043, \dodoi{10.1088/1475-7516/2025/04/043}

\bibitem[{{Faurobert} {et~al.}(2001){Faurobert}, {Arnaud}, {Vigneau}, \& {Frisch}}]{Faurobert2001}
{Faurobert}, M., {Arnaud}, J., {Vigneau}, J., \& {Frisch}, H. 2001, \aap, 378, 627, \dodoi{10.1051/0004-6361:20011263}

\bibitem[{{Faurobert-Scholl} {et~al.}(1995){Faurobert-Scholl}, {Feautrier}, {Machefert}, {Petrovay}, \& {Spielfiedel}}]{Faurobert-Scholl1995}
{Faurobert-Scholl}, M., {Feautrier}, N., {Machefert}, F., {Petrovay}, K., \& {Spielfiedel}, A. 1995, \aap, 298, 289

\bibitem[{{Fontenla} {et~al.}(1993){Fontenla}, {Avrett}, \& {Loeser}}]{Fontenla1993}
{Fontenla}, J.~M., {Avrett}, E.~H., \& {Loeser}, R. 1993, \apj, 406, 319, \dodoi{10.1086/172443}

\bibitem[{{Fraschetti} {et~al.}(2022){Fraschetti}, {Alvarado-G{\'o}mez}, {Drake}, {Cohen}, \& {Garraffo}}]{Fraschetti2022}
{Fraschetti}, F., {Alvarado-G{\'o}mez}, J.~D., {Drake}, J.~J., {Cohen}, O., \& {Garraffo}, C. 2022, \apj, 937, 126, \dodoi{10.3847/1538-4357/ac86d7}

\bibitem[{{Fraschetti} \& {Jokipii}(2011)}]{Fraschetti2011}
{Fraschetti}, F., \& {Jokipii}, J.~R. 2011, \apj, 734, 83, \dodoi{10.1088/0004-637X/734/2/83}

\bibitem[{{Giacalone}(2021)}]{Giacalone2021}
{Giacalone}, J. 2021, \apj, 912, 83, \dodoi{10.3847/1538-4357/abf0b2}

\bibitem[{{Giacalone} \& {Jokipii}(1999)}]{Giacalone1999}
{Giacalone}, J., \& {Jokipii}, J.~R. 1999, \apj, 520, 204, \dodoi{10.1086/307452}

\bibitem[{{Gonz{\'a}lez-Avil{\'e}s} {et~al.}(2018){Gonz{\'a}lez-Avil{\'e}s}, {Guzm{\'a}n}, {Fedun}, {Verth}, {Shelyag}, \& {Regnier}}]{Gonzales2018}
{Gonz{\'a}lez-Avil{\'e}s}, J.~J., {Guzm{\'a}n}, F.~S., {Fedun}, V., {et~al.} 2018, \apj, 856, 176, \dodoi{10.3847/1538-4357/aab36f}

\bibitem[{{Gonz{\'a}lez-Avil{\'e}s} {et~al.}(2021){Gonz{\'a}lez-Avil{\'e}s}, {Murawski}, {Srivastava}, {Zaqarashvili}, \& {Gonz{\'a}lez-Esparza}}]{Gonzales2021}
{Gonz{\'a}lez-Avil{\'e}s}, J.~J., {Murawski}, K., {Srivastava}, A.~K., {Zaqarashvili}, T.~V., \& {Gonz{\'a}lez-Esparza}, J.~A. 2021, \mnras, 505, 50, \dodoi{10.1093/mnras/stab1261}

\bibitem[{{Guti{\'e}rrez} {et~al.}(2022){Guti{\'e}rrez}, {Masip}, \& {Mu{\~n}oz}}]{Gutierrez2022}
{Guti{\'e}rrez}, M., {Masip}, M., \& {Mu{\~n}oz}, S. 2022, \apj, 941, 86, \dodoi{10.3847/1538-4357/aca020}

\bibitem[{{Hudson} {et~al.}(2020){Hudson}, {MacKinnon}, {Szydlarski}, \& {Carlsson}}]{Hudson2020}
{Hudson}, H.~S., {MacKinnon}, A., {Szydlarski}, M., \& {Carlsson}, M. 2020, \mnras, 491, 4852, \dodoi{10.1093/mnras/stz3373}

\bibitem[{{Ishikawa} {et~al.}(2021){Ishikawa}, {Trujillo Bueno}, {del Pino Alem{\'a}n}, {Okamoto}, {McKenzie}, {Auch{\`e}re}, {Kano}, {Song}, {Yoshida}, {Rachmeler}, {Kobayashi}, {Hara}, {Kubo}, {Narukage}, {Sakao}, {Shimizu}, {Suematsu}, {Bethge}, {De Pontieu}, {Sainz Dalda}, {Vigil}, {Winebarger}, {Alsina Ballester}, {Belluzzi}, {{\v{S}}t{\v{e}}p{\'a}n}, {Ramos}, {Carlsson}, \& {Leenaarts}}]{Ishikawa2021}
{Ishikawa}, R., {Trujillo Bueno}, J., {del Pino Alem{\'a}n}, T., {et~al.} 2021, Science Advances, 7, eabe8406, \dodoi{10.1126/sciadv.abe8406}

\bibitem[{{Kafexhiu} {et~al.}(2014){Kafexhiu}, {Aharonian}, {Taylor}, \& {Vila}}]{Kafexhiu2014}
{Kafexhiu}, E., {Aharonian}, F., {Taylor}, A.~M., \& {Vila}, G.~S. 2014, \prd, 90, 123014, \dodoi{10.1103/PhysRevD.90.123014}

\bibitem[{{Kelner} {et~al.}(2006){Kelner}, {Aharonian}, \& {Bugayov}}]{Kelner2006}
{Kelner}, S.~R., {Aharonian}, F.~A., \& {Bugayov}, V.~V. 2006, \prd, 74, 034018, \dodoi{10.1103/PhysRevD.74.034018}

\bibitem[{{Li} {et~al.}(2024{\natexlab{a}}){Li}, {Beacom}, {Griffith}, \& {Peter}}]{Li2024}
{Li}, J.-T., {Beacom}, J.~F., {Griffith}, S., \& {Peter}, A. H.~G. 2024{\natexlab{a}}, \apj, 961, 167, \dodoi{10.3847/1538-4357/ad158f}

\bibitem[{{Li} {et~al.}(2024{\natexlab{b}}){Li}, {Ng}, {Chen}, {Nan}, \& {He}}]{LiZhe2024}
{Li}, Z., {Ng}, K. C.~Y., {Chen}, S., {Nan}, Y., \& {He}, H. 2024{\natexlab{b}}, Chinese Physics C, 48, 045101, \dodoi{10.1088/1674-1137/ad1cda}

\bibitem[{{Linden} {et~al.}(2022){Linden}, {Beacom}, {Peter}, {Buckman}, {Zhou}, \& {Zhu}}]{Linden2022}
{Linden}, T., {Beacom}, J.~F., {Peter}, A. H.~G., {et~al.} 2022, \prd, 105, 063013, \dodoi{10.1103/PhysRevD.105.063013}

\bibitem[{{Linden} {et~al.}(2018){Linden}, {Zhou}, {Beacom}, {Peter}, {Ng}, \& {Tang}}]{Linden2018}
{Linden}, T., {Zhou}, B., {Beacom}, J.~F., {et~al.} 2018, \prl, 121, 131103, \dodoi{10.1103/PhysRevLett.121.131103}

\bibitem[{{Mazziotta} {et~al.}(2020){Mazziotta}, {Luque}, {Di Venere}, {Fass{\`o}}, {Ferrari}, {Loparco}, {Sala}, \& {Serini}}]{Mazziotta2020}
{Mazziotta}, M.~N., {Luque}, P. D. L.~T., {Di Venere}, L., {et~al.} 2020, \prd, 101, 083011, \dodoi{10.1103/PhysRevD.101.083011}

\bibitem[{{Mignone} {et~al.}(2007){Mignone}, {Bodo}, {Massaglia}, {Matsakos}, {Tesileanu}, {Zanni}, \& {Ferrari}}]{Mignone2007}
{Mignone}, A., {Bodo}, G., {Massaglia}, S., {et~al.} 2007, \apjs, 170, 228, \dodoi{10.1086/513316}

\bibitem[{{Mignone} {et~al.}(2018){Mignone}, {Bodo}, {Vaidya}, \& {Mattia}}]{Mignone2018}
{Mignone}, A., {Bodo}, G., {Vaidya}, B., \& {Mattia}, G. 2018, \apj, 859, 13, \dodoi{10.3847/1538-4357/aabccd}

\bibitem[{{Mignone} {et~al.}(2012){Mignone}, {Zanni}, {Tzeferacos}, {van Straalen}, {Colella}, \& {Bodo}}]{Mignone2012}
{Mignone}, A., {Zanni}, C., {Tzeferacos}, P., {et~al.} 2012, \apjs, 198, 7, \dodoi{10.1088/0067-0049/198/1/7}

\bibitem[{{Morton} {et~al.}(2023){Morton}, {Sharma}, {Tajfirouze}, \& {Miriyala}}]{Morton2023}
{Morton}, R.~J., {Sharma}, R., {Tajfirouze}, E., \& {Miriyala}, H. 2023, Reviews of Modern Plasma Physics, 7, 17, \dodoi{10.1007/s41614-023-00118-3}

\bibitem[{{Ng} {et~al.}(2016){Ng}, {Beacom}, {Peter}, \& {Rott}}]{Ng2016}
{Ng}, K. C.~Y., {Beacom}, J.~F., {Peter}, A. H.~G., \& {Rott}, C. 2016, \prd, 94, 023004, \dodoi{10.1103/PhysRevD.94.023004}

\bibitem[{{Ng} {et~al.}(2024){Ng}, {Hillier}, \& {Ando}}]{Ng2024}
{Ng}, K. C.~Y., {Hillier}, A., \& {Ando}, S. 2024, arXiv e-prints, arXiv:2405.17549, \dodoi{10.48550/arXiv.2405.17549}

\bibitem[{{Puzzoni} {et~al.}(2024){Puzzoni}, {Fraschetti}, {K{\'o}ta}, \& {Giacalone}}]{Puzzoni2024}
{Puzzoni}, E., {Fraschetti}, F., {K{\'o}ta}, J., \& {Giacalone}, J. 2024, \apj, 973, 118, \dodoi{10.3847/1538-4357/ad65ea}

\bibitem[{{Rankin}(2024)}]{Rankin2024}
{Rankin}, J.~S. 2024, in AGU Fall Meeting Abstracts, Vol. 2024, SH21C--2832

\bibitem[{{Reale}(2014)}]{Reale:14}
{Reale}, F. 2014, Living Reviews in Solar Physics, 11, 4, \dodoi{10.12942/lrsp-2014-4}

\bibitem[{{Rempel}(2014)}]{Rempel2014}
{Rempel}, M. 2014, \apj, 789, 132, \dodoi{10.1088/0004-637X/789/2/132}

\bibitem[{{Schad} {et~al.}(2024){Schad}, {Petrie}, {Kuhn}, {Fehlmann}, {Rimmele}, {Tritschler}, {Woeger}, {Scholl}, {Williams}, {Harrington}, {Paraschiv}, \& {Szente}}]{Schad2024}
{Schad}, T.~A., {Petrie}, G., {Kuhn}, J., {et~al.} 2024, Science Advances, 10, eadq1604, \dodoi{10.1126/sciadv.adq1604}

\bibitem[{{Seckel} {et~al.}(1991){Seckel}, {Stanev}, \& {Gaisser}}]{Seckel1991}
{Seckel}, D., {Stanev}, T., \& {Gaisser}, T.~K. 1991, \apj, 382, 652, \dodoi{10.1086/170753}

\bibitem[{{Shchukina} \& {Trujillo Bueno}(2011)}]{Shchukina2011}
{Shchukina}, N., \& {Trujillo Bueno}, J. 2011, \apjl, 731, L21, \dodoi{10.1088/2041-8205/731/1/L21}

\bibitem[{{Steiner} {et~al.}(2008){Steiner}, {Rezaei}, {Schaffenberger}, \& {Wedemeyer-B{\"o}hm}}]{Steiner.etal:08}
{Steiner}, O., {Rezaei}, R., {Schaffenberger}, W., \& {Wedemeyer-B{\"o}hm}, S. 2008, \apjl, 680, L85, \dodoi{10.1086/589740}

\bibitem[{{Stenflo}(2013)}]{Stenflo2013}
{Stenflo}, J.~O. 2013, \aap, 555, A132, \dodoi{10.1051/0004-6361/201321608}

\bibitem[{{Tang} {et~al.}(2018){Tang}, {Ng}, {Linden}, {Zhou}, {Beacom}, \& {Peter}}]{Tang2018}
{Tang}, Q.-W., {Ng}, K. C.~Y., {Linden}, T., {et~al.} 2018, \prd, 98, 063019, \dodoi{10.1103/PhysRevD.98.063019}

\bibitem[{{Trujillo Bueno} {et~al.}(2004){Trujillo Bueno}, {Shchukina}, \& {Asensio Ramos}}]{Trujillo2004}
{Trujillo Bueno}, J., {Shchukina}, N., \& {Asensio Ramos}, A. 2004, \nat, 430, 326, \dodoi{10.1038/nature02669}

\bibitem[{{Zhou} {et~al.}(2017){Zhou}, {Ng}, {Beacom}, \& {Peter}}]{Zhou2017}
{Zhou}, B., {Ng}, K. C.~Y., {Beacom}, J.~F., \& {Peter}, A. H.~G. 2017, \prd, 96, 023015, \dodoi{10.1103/PhysRevD.96.023015}

\end{thebibliography}
\bibliographystyle{aasjournal}



\end{document}